\documentclass[lettersize,journal]{IEEEtran}
\usepackage{amsmath,amsfonts}
\usepackage{algorithm}
\usepackage{algpseudocode}
\usepackage{amsmath}
\usepackage{array}
\usepackage[caption=false,font=footnotesize,labelfont=rm,textfont=rm]{subfig}
\usepackage{textcomp}
\usepackage{stfloats}
\usepackage{url}
\usepackage{verbatim}
\usepackage{graphicx}
\usepackage{cite}
\usepackage{xcolor}
\usepackage{hhline}
\usepackage{booktabs}
\usepackage{tabularx}
\usepackage{float}
\usepackage{amsmath}
\usepackage{multirow}
\usepackage{makecell}
\usepackage{cite}
\usepackage{colortbl}
\definecolor{lightgray}{gray}{0.95}

\usepackage{hyperref}
\begin{document}
	
	\title{SAR Despeckling via Region-Aware Sparse Representation and Statistical Noise Approximation}
	
	\author{
		Xuran Hu,
		Mingzhe Zhu,
		Djordje Stankovi\'c,~\IEEEmembership{Member,~IEEE},\\
		Zhenpeng Feng,
		Yifang Ban,~\IEEEmembership{Senior Member,~IEEE},
		and Ljubi\v{s}a Stankovi\'c,~\IEEEmembership{Fellow,~IEEE}

		\thanks{X. Hu, Z. Feng and M. Zhu are with School of Electronic Engineering, Xidian University, Xi’an, China. Email: XuRanHu@stu.xidian.edu.cn, zpfeng\_1@stu.xidian.edu.cn, zhumz@mail.xidian.edu.cn.}
		\thanks{X. Hu and M. Zhu are also with Kunshan Innovation Institute of Xidian University, School of Electronic Engineering, Xidian University, Xi'an , China}
		\thanks{L. Stankovi\'c and Djordje Stankovi\'c are with the EE Department of the University of Montenegro, Podgorica, Montenegro. Email: ljubisa@ucg.ac.me, dstankovic@ucg.ac.me.}
		\thanks{X. Hu and Y. Ban are with Division of Geoinformatics, KTH Royal Institute of Technology. Email: xuranh@kth.se, yifang@kth.se}
		\thanks{Corresponding author: Mingzhe Zhu}}
	
	\markboth{Journal of \LaTeX\ Class Files,~Vol.~14, No.~8, August~2021}%
	{Shell \MakeLowercase{\textit{et al.}}: A Sample Article Using IEEEtran.cls for IEEE Journals}
	
	\IEEEpubid{0000--0000/00\$00.00~\copyright~2021 IEEE}
	
	\maketitle

\begin{abstract}
Synthetic Aperture Radar (SAR) imagery are widely utilized in remote sensing due to their all-weather, all-day imaging capabilities. However, SAR images are highly susceptible to noise, particularly speckle noise, caused by the coherent imaging process, which severely degrades image quality. This has driven increasing research interest in SAR despeckling. Sparse representation-based methods have been extensively applied in natural image processing, yet SAR despeckling requires addressing non-Gaussian assumption and ensuring sparsity in the transform domain. In this work, we propose a simple, intuitive, and efficient SAR despeckling approach grounded in compressive sensing theory. By applying Log-Yeo-Johnson transformation, we convert gamma-distributed noise into an approximate Gaussian distribution to noise sparse assumption. The method incorporates noise and sparsity priors, leveraging a non-local sparse representation through auxiliary matrices: one capturing varying noise characteristics across regions and the other encoding adaptive sparsity information. Extensive experiments validate the effectiveness of our method.
\end{abstract}


\section{Introduction}
\IEEEPARstart{S}{ynthetic} Aperture Radar (SAR) imagery is widely utilized in earth observation \cite{11075901}, electronic reconnaissance \cite{hu2024manifold}, and disaster monitoring \cite{zhao2025radarsat} due to its all-weather, all-time capabilities. However, due to its coherent side-looking imaging characteristics, these images often contain substantial speckle noise, which significantly reduces image quality and adversely affects downstream detection and recognition tasks. This has led to growing interest in SAR despeckling.

Noise-free ground truth samples are unavailable for SAR images, which introduces new challenges for SAR despeckling. Common despeckling methods include traditional filtering, transform-based methods, prior-based approaches \cite{baraha2023speckle}, and deep learning methods \cite{wang2017sar}. This paper focuses on prior-based methods, which estimate noise-free images by leveraging various types of prior information. Specifically, these approaches build maximum a posteriori (MAP) estimation models incorporating different priors and optimize the loss function to achieve optimal despeckling results under the given observations. Typical priors include total variation \cite{AFONSO2015101}, sparsity \cite{baraha2023speckle}, low-rank \cite{10032489}, and denoiser-induced \cite{bai2019new} priors.

Compared to optical image denoising, SAR despeckling faces two primary challenges. The first is how to adaptively recover a noise-free image without access to a noise-free ground truth. Recently, several deep learning-based methods \cite{hu2024sar} have been proposed, achieving promising results. However, most of these rely on supervised learning \cite{9099060}, which requires extensive effort to construct synthetic datasets and carries the risk of overfitting, reducing robustness and adaptability of the model. To alleviate the dependence on clean references, recent works have explored self-supervised paradigms based on Deep Image Prior~\cite{albisani2025self}, masked Transformer pretext tasks~\cite{lin2025speckle2self}, and multi-channel projections~\cite{denis2025just}, which extract supervision directly from the noisy observations. In parallel, efficient conditional diffusion models~\cite{guo2025efficient} have been proposed to mitigate the iterative sampling cost of standard DDPMs. Despite these advances, learning-based methods remain bounded by their training distributions and exhibit substantial inference overhead in practical deployment. Traditional filtering and transform-based methods also encounter this problem, as their performance is highly sensitive to settings of hyperparameter under varying noise levels. Prior-based methods address this issue by introducing diverse prior information to guide denoising, yet the adaptive nature of these methods remains limited due to the complexity of SAR noise types and intricate texture details, which may not align with low-rank \cite{xu2023edge} or sparsity assumptions \cite{stankovic2022image}, leading to non-convex optimization.

The second challenge in SAR despeckling is performing denoising under gamma noise priors. SAR images contain varying levels of speckle noise. Thus, directly applying gamma priors often fails to construct a convex optimization model, resulting in local optima. Some methods mitigate this by converting multiplicative gamma noise into additive Gaussian noise for subsequent processing \cite{ma2024despeckling}. However, commonly used transformations like Log-Yeo-Johnson or Box-Cox only approximate Gaussian distribution, undermining the reliability of Gaussian-based models. A very recent line of work has attempted to overcome this limitation by embedding the Log-Yeo--Johnson transformation within learning-based frameworks. For instance, \cite{heo2026self} couples it with self-supervised score-based diffusion to model the noise residual in a learned manifold, \IEEEpubidadjcol More broadly, frequency-adaptive deep models~\cite{ma2025towards} have been proposed to address the spatially heterogeneous nature of speckle. These approaches achieve strong performance but require large training corpora and remain computationally intensive at both training and inference stages, motivating the training-free, model-based framework explored in this work.

Sparse representation-based methods for despeckling originated from compressive sensing theory \cite{elad2006image}, aiming to maximize sparsity in transform domain of image to remove high-frequency noise. By defining basis matrix, sparse representations of the image can be obtained. The compressive sensing model reformulates image denoising as a Lasso optimization problem with $l_1$ regularization, solved using methods like Alternating Direction Method of Multipliers (ADMM) \cite{ouyang2013stochastic} or Iterative Shrinkage-Thresholding Algorithm (ISTA). In the context of SAR despeckling, sparsity-guided methods ensure the transform domain adheres to sparsity assumptions, achieving optimal noise-free estimation through well-chosen priors. Various approaches have enhanced performance by introducing weighted regularization, dictionary learning \cite{cui2024one}, or hybrid models \cite{bo2024novel} combining low-rank and other priors.

This paper proposes a simple, intuitive, and efficient SAR despeckling approach based on sparse representation and compressive sensing. We first apply Log-Yeo-Johnson transformation to convert the multiplicative gamma noise in SAR images into approximate additive Gaussian noise. Following this, a non-local sparse prior is introduced to account for noise variation across different SAR regions, allowing noise distributions of image patch to better align with Gaussian assumptions, ensuring model effectiveness. Furthermore, we identify similar features using Euclidean distance, establishing a non-local compressive sensing model. Two auxiliary matrices are introduced to better characterize noise properties across image regions and to align with sparse priors.

The main contributions of this work are as follows:

\begin{enumerate}
	\item We propose a sparse representation-guided SAR despeckling method that achieves adaptive, high-performance despeckling by fully incorporating SAR noise and sparsity priors.
	
	\item We introduce Log-Yeo-Johnson transformation to preprocess SAR images and incorporate a non-local sparse model in the subsequent MAP estimation, effectively accounting for variations across image patches, enhancing adherence to Gaussian noise assumptions.
	
	\item We propose a fully training-free SAR despeckling framework that operates without any clean reference image. Owing to its model-based formulation, the method adapts automatically across arbitrary spatial resolutions, frequency bands, and noise levels, as validated on eight datasets with diverse sensors and acquisition conditions.
\end{enumerate}

\section{Related Works}

\subsection{SAR Despeckling via Learning}

Recently, deep learning-based methods have become a major focus for SAR image despeckling, evolving from early CNN-based models to more advanced generative, self-supervised, and diffusion-based approaches.

\cite{kevala2024nbdnet} introduced attention modules into a CNN architecture to better preserve structural and textural information, but its reliance on synthetic noisy–clean pairs limits domain adaptability. \cite{hu2024sar} proposed a region-guided diffusion model that improves multi-scale despeckling consistency, though it remains computationally intensive. \cite{pan2024sar} incorporated the Swin Transformer into a diffusion framework to enhance long-range texture modeling, at the cost of increased memory and inference requirements. \cite{takao2025md} combined despeckling and super-resolution using a multi-task flow-based model, but at the expense of greater inference complexity. \cite{fang2024contrastive} used contrastive learning to enhance speckle–texture separation without requiring labels, yet it is sensitive to sample selection. \cite{perera2022sar} employed an overcomplete dual-branch CNN to model both local and global features, balancing structure and detail.

Several self-supervised or blind-spot models also exist. SAR2SAR \cite{dalsasso2021sar2sar} is a representative self-supervised despeckling method. Its core idea is inspired by Noise2Noise framework, treating multi-temporal SAR images of the same area as independent noise realizations for training, thus utilizing real noise statistical relationships instead of synthetic clean references for supervision. By pairing multiple SAR images of the same region, the method constrains the network to preserve stable terrain structures while suppressing random speckle noise. This greatly reduces the reliance on ideal noise-free images and achieves performance superior to traditional filters and supervised CNNs on multi-source datasets such as Sentinel-1 and TerraSAR-X. Speckle2Void \cite{molini2021speckle2void} removes the need for clean reference images but may oversmooth fine textures. \cite{saha2024despeckling} introduced gradient constraints in CNN+GAN to better preserve edges, while PolMERLIN \cite{kato2024polmerlin} uses polarimetric information for improved multi-channel despeckling. The latest advances have pushed deep despeckling toward self-supervision, efficiency, and statistical interpretability. S3DIP~\cite{albisani2025self} extends the Deep Image Prior framework with a learnable speckle matrix and a differentiable histogram loss, enabling fully training-free per-image despeckling. Speckle2Self~\cite{lin2025speckle2self} formulates despeckling as a masked-pixel estimation problem on a Transformer backbone, achieving performance comparable to supervised methods without requiring multi-temporal data. \cite{denis2025just} generalizes the MERLIN self-supervised 
strategy to multi-channel SAR via projection-based decomposition, restoring statistical independence between sub-channels. 
\cite{heo2026self} establishes a connection between the Log-Yeo--Johnson transformation and score-based diffusion, leveraging Tweedie's formula for clean-reference-free training. Hybrid algorithm-network designs, such as the diffusion-equation-stabilized unrolling network in~\cite{ran2025tunable}, the frequency-adaptive heterogeneous learner in~\cite{ma2025towards}, and the efficient conditional diffusion model, further address the trade-off between speckle suppression and computational cost. Beyond the despeckling literature, \cite{gao2025diffusion} also demonstrates that diffusion priors can serve as plug-in regularizers in broader SAR inverse problems. Collectively, these methods define the current frontier of learning-based despeckling, yet each still inherits the fundamental limitations of data-driven models: dependence on representative training data, sensitivity to distribution shift across sensors, and lack of deterministic guarantees, which sustain the case for principled model-based alternatives.

Although deep learning methods have achieved notable progress, they still face challenges: the lack of truly clean SAR references limits supervised training; self-supervised or contrastive approaches often require assumptions (e.g., channel independence) that do not always hold; overfitting remains an issue under diverse imaging conditions; and the black-box nature of deep models reduces interpretability and trustworthiness, which restricts their usage in sensitive applications compared to traditional methods.

\subsection{SAR Despeckling via Sparse Representation}

Sparse representation, also known as compressive sensing, enables efficient image reconstruction with a sampling rate significantly lower than that required by Nyquist sampling theorem. Furthermore, researchers have found that compressive sensing can be leveraged for efficient image denoising. Sparse representation-based denoising primarily emphasizes sparsity of the transform domain in the objective function (i.e., the L1 regularization term) while relaxing the reconstruction error constraint, thereby transforming the reconstruction model into a denoising model.

In recent years, an increasing number of methods have introduced the gamma noise model and the concept of nonlocal sparsity, aiming to better capture the noise priors and characteristics in different regions to guide SAR image despeckling. \cite{lu2016sar} proposed a structural sparse representation-based method for SAR despeckling, designing separate sparse models for different regions. \cite{zhang2020learning} used least squares estimation to recover noise-free images by modeling the residuals between the noisy and clean images via a sparse representation framework. \cite{LIU2017174} proposed a classification-based SAR despeckling approach, where different sparse representation models are built for different categories of image patches. Liu et al. Although achieve acceptable performance, most of these models do not fully account for the unique noise characteristics of SAR images, thus achieving only limited performance. \cite{9484779} and \cite{xu2018trilateral} proposed weighted nonlocal sparse models for SAR image despeckling; nevertheless, these approaches are based on Gaussian priors, and since gamma noise does not comply with this assumption, the models fail to achieve optimal despeckling performance.

To address this limitation, some methods have directly formulated maximum a posteriori estimation models based on gamma or Rayleigh distributions. For instance, \cite{liu2017over} presented an overcomplete dictionary-based sparse representation method. In this approach, a mean filter is first applied to suppress speckle noise while preserving structural information, followed by learning a dictionary to estimate the noise-free image. The mean filtering step alters the noise characteristics of SAR images, potentially making the noise better satisfy the assumptions of the MAP framework. \cite{8629005} employed a group sparse representation model solved via the split Bregman iterative algorithm, which preserves image structures and achieves improved reconstruction performance. \cite{10032489} directly modeled multiplicative noise, incorporating Rayleigh distribution properties and nonlocal sparsity to achieve effective SAR despeckling.

While these methods partially address the limitations of prior assumptions in MAP estimation, they also introduce new challenges. Models based on multiplicative noise are often non-convex, making it difficult to approach the global optimum. Moreover, SAR image scenes are typically more complex, with varying noise intensities and characteristics across different regions. Traditional nonlocal approaches mainly focus on searching for similar patches but do not dynamically adjust the learning strategy according to the noise intensity. In parallel, several latest works have revisited model-based despeckling from a statistical perspective: \cite{zafari2025bayesian} proposed a Bayesian despeckling framework that explicitly models structured signal priors beyond local stationarity assumptions, while diffusion priors have been used as regularizers in broader SAR inverse problems with incomplete measurements~\cite{gao2025diffusion}. These developments indicate that sparse-representation and Bayesian estimation remain active foundations even in the deep-learning era, particularly for 
scenarios demanding interpretability, deterministic behavior, and zero training overhead. To tackle these issues, this paper proposes a sparse representation model that integrates region awareness and noise transformation, enabling effective SAR despeckling.

\section{Methodology}

\subsection{Overview of Proposed Model}

To address the challenges posed by the multiplicative gamma noise and spatially varying characteristics of SAR images, we propose a region-aware sparse representation framework that combines statistical noise transformation with non-local sparse modeling. Specifically, the method first applies a Log-Yeo-Johnson transformation to approximate the gamma-distributed noise as additive Gaussian noise, enabling the use of sparse priors. Then, we perform non-local patch grouping based on Euclidean similarity to construct a compressive sensing model guided by auxiliary matrices. These matrices encode local noise levels and adaptive sparsity priors, ensuring region-sensitive denoising. The final sparse coding is formulated as a weighted Lasso problem, solved via an ADMM-based optimization. This framework enables effective despeckling across different SAR scenes by integrating noise-aware modeling, structural sparsity, and patch-level adaptivity in a unified way.

\subsection{Log-Yeo-Johnson Transformation for Noise Conversion}

SAR images typically contain speckle noise. Assuming a SAR image $y$, it can be represented as: 

\begin{equation}
	y=x \cdot n
\end{equation} where $x$ is the noise-free ground-truth image, and $n$ is multiplicative gamma noise. The probability density function of gamma noise $n$ is defined as: 

\begin{equation}
	p(n)=\frac{L^L n^{L-1} e^{-L n}}{\Gamma(L)}
\end{equation} where $L$ is equivalent number of looks (ENL), and $\Gamma(\cdot)$ denotes the gamma function.

To leverage noise characteristics effectively for constructing a compressive sensing model, we adopt Log-Yeo-Johnson transformation \cite{ma2024despeckling} to approximate the gamma noise to a Gaussian distribution, based on prior research. First, a logarithmic transformation converts the multiplicative noise into additive noise: 

\begin{equation}
	\log (y)=\log (x)+\log (n)
\end{equation}

Following this, we apply the Yeo-Johnson transformation to further process the image, converting it into an approximate Gaussian distribution. The Yeo-Johnson transformation is defined as follows:

\begin{equation}
	\operatorname{Yeo}(x)= \begin{cases}\frac{(x+1)^\lambda-1}{\lambda}, & x \geq 0, \lambda \neq 0 \\ \log (x+1), & x \geq 0, \lambda=0 \\ -\frac{(-x+1)^{2-\lambda}-1}{2-\lambda}, & x<0, \lambda \neq 2 \\ -\log (-x+1), & x < 0, \lambda=2\end{cases}
\end{equation} where $\lambda$ is a shape parameter selected according to the data properties, here optimized by minimizing kurtosis skewness.

We later revert the transformed image back to the spatial domain using the inverse Log-Yeo-Johnson transformation, which is defined as:

\begin{equation}
	\text { Yeo }^{-1}(x)= \begin{cases}(\lambda x+1)^{\frac{1}{\lambda}}-1, & x \geq 0, \lambda \neq 0 \\ e^x-1, & x \geq 0, \lambda=0 \\ -[(\lambda-2) x+1]^{\frac{1}{2-\lambda}}, & x<0, \lambda \neq 2 \\ -e^{-x}+1, & x < 0, \lambda=2\end{cases}
\end{equation}

\begin{figure*}[t]
	\centering
	\includegraphics[width=\textwidth]{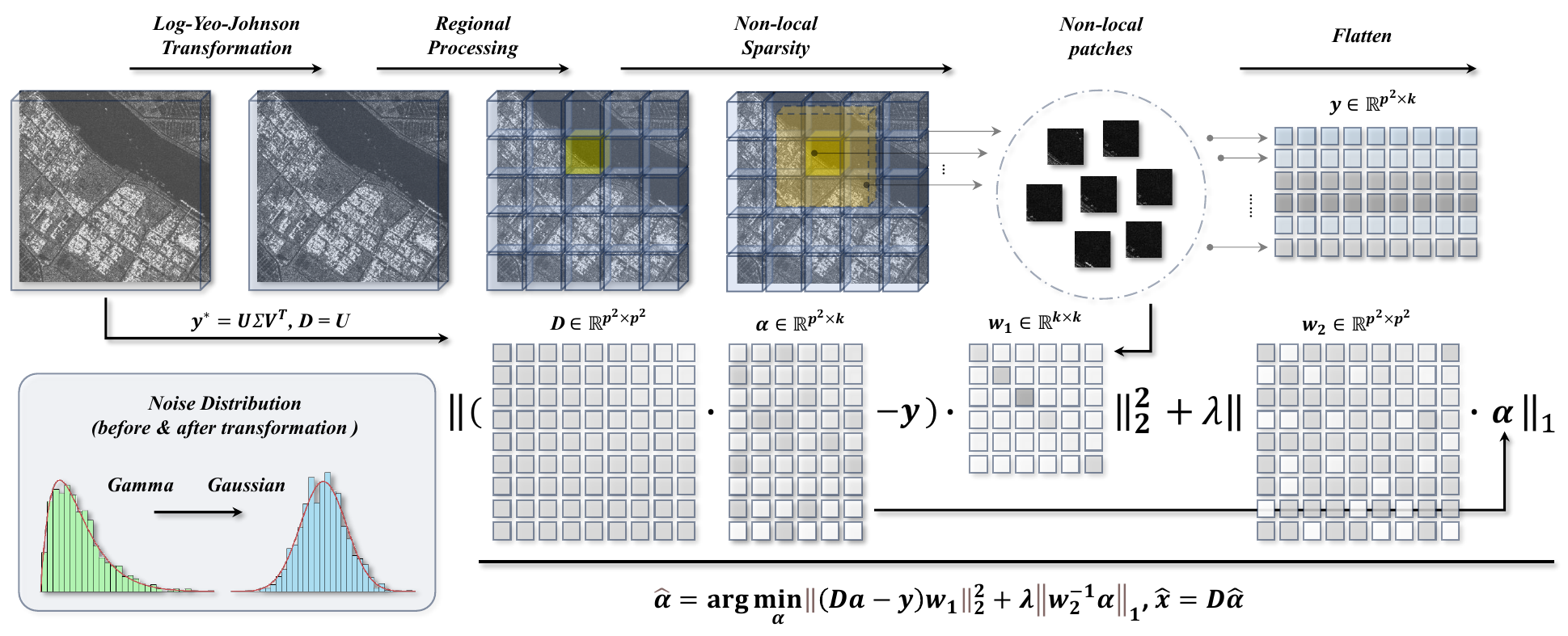}
	\caption{Flowchart of our proposed method: (1) Gamma noise is approximated as Gaussian through noise transformation; (2) non-local sparse modeling is performed, guided by auxiliary matrices $w_1$ and $w_2$ derived from noise level and sparsity.}
	\label{fig_1}
\end{figure*}

\subsection{Non-local Sparse-Guided SAR Despeckling}

According to compressive sensing theory, each image patch $x \in \mathbb{R}^{p \times p}$ can be sparsely represented by a dictionary, which may be derived using singular value decomposition \cite{aharon2006k} or bases from certain transform domains, such as discrete cosine transform (DCT) or wavelet transform. The mathematical formulation of compressive sensing is as follows:

\begin{equation}
	\hat{\alpha}=\arg \min _\alpha\|\alpha\|_0 \text { s.t. } D \alpha=x
\end{equation} where $\|\cdot\|_0$ denotes the $l_0$-norm. Here, $\|\hat{\alpha}\|_0$ is significantly smaller than dimension $p \times p$. Intuitively, the image can be represented as a linear combination of dictionary atoms.

For image denoising \cite{elad2006image}, given a noisy image $y$, the equation for $x$ is modified as:

\begin{equation}
	\hat{\alpha}=\arg \min _\alpha\|\alpha\|_0 \text { s.t. }\|D \alpha-y\|_2^2 \leq t(\epsilon, \sigma)
\end{equation} where $t(\epsilon, \sigma)$ is a function of $\epsilon$ (the permissible error) and $\sigma$ (the noise level). Since we previously transformed the gamma noise in SAR images to an approximate Gaussian distribution, $\sigma$ now represents the standard deviation of the Gaussian noise.

To facilitate optimization, we replace the $l_0$-norm with the $l_1$-norm, transforming the problem into a Lasso problem.

Most SAR despeckling methods operate via block processing; here, we achieve SAR despeckling by leveraging the sparse representation of non-local image patches. For a patch $y_0 \in \mathbb{R}^{p \times p}$, we identify $k$ similar patches by calculating their Euclidean distances $d\left(y_0, y_i\right)=\left\|y_0-y_i\right\|^2$, stacking them by similarity into a noise patch matrix $y \in \mathbb{R}^{p \times p \times k}$, where the most similar patch is the original patch itself. We then substitute this matrix into equation $x$ to compute its sparse representation $\alpha$:

\begin{equation}
	\hat{\alpha}=\arg \min _\alpha\|D \alpha-y\|_2^2+c\|\alpha\|_1
\end{equation} where $c$ is a regularization parameter. This work applies the Alternating Direction Method of Multipliers (ADMM) to solve the Lasso problem, obtaining the optimal sparse representation $\hat{\alpha}$, and producing the despeckled image $\hat{x}=D \hat{\alpha}$.

For the sparse basis $D$, we employ singular value decomposition (SVD) to derive it. Similar to other compressive sensing techniques, once $D$ is determined, subsequent Lasso solutions remain unchanged:

\begin{algorithm}[t]
	\caption{SAR Despeckling via Non-local Sparsity}
	\label{alg:sar_despeckling}
	\begin{algorithmic}[1]
		\State \textbf{Input:} Noisy SAR image $y$, dictionary $D$, parameters $\lambda$, $c$, $\eta$, $K_{\text{max}}$, $T_{\max}$, $\text{Tol}$
		\State \textbf{Output:} Denoised image $\hat{x}$
		
		\vspace{0.3em}
		\State Apply log transform: $\log(y) = \log(x) + \log(n)$
		\State Apply Yeo-Johnson: $y_{\text{trans}} = \text{transform}(y, \lambda)$
		\State Global noise init: $\hat{\sigma}_0 \gets \kappa \cdot \mathrm{median}(|c_D|)$, $\sigma_{\min} \gets \eta \hat{\sigma}_0$
		
		\vspace{0.3em}
		\For{each patch $y_0$ in $y_{\text{trans}}$}
		\State Find $K$ similar patches: $d(y_0, y_i) = \|y_0 - y_i\|^2$
		\State Stack into matrix $Y \in \mathbb{R}^{p^2 \times K}$
		\EndFor
		
		\vspace{0.3em}
		\State Initialize: $\alpha_0 = Z_0 = \Delta_0 = 0$, $\rho_0 > 0$, $t = 0$
		\State Initialize weights: $w_1 = \hat{\sigma}_0^{-1} I_K$,\quad $w_2 = \Sigma^{-1}$
		
		\While{$t < T_{\max}$ \textbf{and} not converged}
		
		\vspace{0.3em}
		\State $
		\begin{aligned}
			\alpha_{t+1} = & \arg\min_{\alpha} \|(D\alpha - Y)w_1\|_F^2 \\
			& + \frac{\rho_t}{2} \|w_2 \alpha - Z_t + \rho_t^{-1} \Delta_t\|_F^2
		\end{aligned}$
		
		\vspace{0.3em}
		\State $
		\begin{aligned}
			(D^\top w_1^\top w_1 D &  + \tfrac{\rho_t}{2} w_2^\top w_2) \alpha_{t+1} = \\ 
			& D^\top w_1^\top w_1 Y + \tfrac{\rho_t}{2} w_2^\top (Z_t - \rho_t^{-1} \Delta_t)
		\end{aligned}$
		
		\vspace{0.3em}
		\State $r_k^{(t+1)} \gets y_k - D\alpha_{k,t+1}$
		\State $\hat{\sigma}_k^{(t+1)} \gets \max\!\big(\mathrm{MAD}(r_k^{(t+1)}),\; \sigma_{\min}\big)$
		\State $w_1 \gets \mathrm{diag}\big(1/\hat{\sigma}_1^{(t+1)},\dots,1/\hat{\sigma}_K^{(t+1)}\big)$
		
		\vspace{0.3em}
		\State $Z_{t+1} = S_{c/\rho_t}(w_2 \alpha_{t+1} + \rho_t^{-1} \Delta_t)$
		\State $\Delta_{t+1} = \Delta_t + \rho_t (w_2 \alpha_{t+1} - Z_{t+1})$
		\State $\rho_{t+1} = \mu \rho_t$
		
		\vspace{0.3em}
		\If{$\|w_2 \alpha_{t+1} - Z_{t+1}\|_F \leq \text{Tol}$}
		\State converged $\gets$ \textbf{true}
		\EndIf
		\State $t \gets t + 1$
		\EndWhile
		\State $\hat{\alpha} = \alpha_{t}$
		
		\vspace{0.3em}
		\State $\hat{x}_{\text{trans}} \gets D\hat{\alpha}$
		\State Apply inverse Yeo-Johnson: $\text{Yeo}^{-1}(\hat{x}_{\text{trans}})$
		\State $\hat{x} \gets \exp(\text{Yeo}^{-1}(\hat{x}_{\text{trans}})) - 1$
		
	\end{algorithmic}
\end{algorithm}

\subsection{Sparsity-guided Posterior Estimation}

To effectively utilize the noise statistical properties after Log-Yeo-Johnson transformation, as well as the sparsity in the transform domain, we introduce two matrices, $w_1$ and $w_2$, to incorporate these information. Similar to typical sparse representation methods, we apply maximum a posteriori estimation to determine $w_1$ and $w_2$. Given the observed noisy image $y$, our objective is to restore the noise-free image $x$, which is equivalent to solving for the sparse representation $\alpha$ as follows:

\begin{equation}
	\hat{\alpha}=\arg \max _\alpha P(\alpha \mid y)
\end{equation}

Following Bayes’ theorem, the posterior probability $P(\alpha \mid y)$ can be decomposed as:

\begin{equation}
	P(\alpha \mid y) \propto P(y \mid \alpha) \cdot P(\alpha)
\end{equation} where $P(y \mid \alpha)$ is the likelihood function of the noisy image $y$ given the sparse representation $\alpha$, reflecting the statistical properties of the noise, and $P(\alpha)$ is the prior probability of the sparse representation, capturing the sparsity of the image.

\begin{equation}
	\begin{gathered}
		y=U \Sigma V^T \\
		D=U
	\end{gathered}
\end{equation}

\noindent \textbf{Modeling the Likelihood Function: }Using Log-Yeo-Johnson transformation, gamma noise in SAR image is approximated as Gaussian noise. We assume that noise in each image patch is independently and identically distributed Gaussian noise. For each patch $y_k$:

\begin{equation}
	P(y \mid \alpha)=\prod_{k=1}^K\left(\pi \sigma_k\right)^{-p^2} e^{-\sigma_k^{-2}\left\|y_k-D \alpha_k\right\|_2^2}
\end{equation} where $y_k$ and $\alpha_k$ denote the $k$-th column of matrices $y$ and $\alpha$, respectively, and $\sigma_k$ is the standard deviation of the noise in patch $y_k$. This probability model captures the variation in noise across patches.

\noindent \textbf{Modeling the Prior: }According to compressive sensing theory, sparse representation $\alpha$ has only a few non-zero atoms, allowing efficient representation of key features. Given the sharp peak characteristic of Laplace distribution's probability density function, we adopt this distribution as our sparsity prior. Assuming each element in matrix $\alpha$ is independently and identically distributed, the distribution for each $\alpha_i^{k}$ follows:

\begin{equation}
	P\left(\alpha_{i k}\right)=\frac{1}{2 S_i} e^{\left(-\frac{\left|\alpha_{i k}\right|}{S_i}\right)}
\end{equation} where $S_i$ is a scale parameter of the sparsity prior, corresponding to the $i$-th atom in dictionary $D$. Specifically, $S_i$ governs the sparsity of each coefficient $\alpha_i^{k}$: for important features in the dictionary (i.e., those with larger singular values), $S_i$ is larger, allowing for higher non-zero values of $\alpha_{ik}$, and vice versa. The values of $S_i$ are derived from singular value decomposition (SVD) and indicate the importance of each feature.

%
%
%
%
%
%
%
%
%
%

Given that each $\alpha_i^{k}$ in $\alpha$ is independently and identically distributed, $P(\alpha)$ can be expressed as a joint probability:

\begin{equation}
	P(\alpha)=\prod_{k=1}^K \prod_{i=1}^{p^2} \frac{1}{2 S_i} e^{\left(-\frac{\left|\alpha_{i k}\right|}{S_i}\right)}
\end{equation}

\noindent \textbf{Maximum A Posteriori Estimation and Objective Function: }Taking the logarithm of $P(\alpha \mid y)$, the objective function becomes:

\begin{equation}
	\begin{aligned}
		\hat{\alpha} 
		&= \arg\max_{\alpha}\, \ln P(\alpha \mid y) \\
		&= \arg\max_{\alpha}\, \big\{\ln P(y \mid \alpha) + \ln P(\alpha)\big\} \\
		&= \arg\max_{\alpha}\, \sum_{k=1}^{K}\!\left[-\frac{p^2}{2}\ln\!\big(2\pi\sigma_k^2\big) 
		- \frac{1}{2\sigma_k^2}\lVert y_k - D\alpha_k \rVert_2^2\right] \\
		&\quad + \sum_{k=1}^{K}\sum_{i=1}^{p^2}\!\Big[-\ln(2 S_i) - S_i^{-1}\,|\alpha_{ik}|\Big].
	\end{aligned}
	\label{eq:map}
\end{equation}

Reformulating further, we can get:

\begin{equation}
	\hat{\alpha} 
	= \arg\min_{\alpha}\, 
	\sum_{k=1}^{K} \frac{1}{2\sigma_k^2}\,\lVert y_k - D\alpha_k \rVert_2^2 
	+ \sum_{k=1}^{K}\sum_{i=1}^{p^2} S_i^{-1}\,|\alpha_{ik}|.
	\label{eq:map-simplified}
\end{equation}

Letting $w_1 = \mathrm{diag}(\sigma_1^{-1}, \dots, \sigma_K^{-1})$ and $w_2 = \Sigma^{-1}$, and absorbing the constant factor $1/2$ into the regularization weight $c$, \eqref{eq:map-simplified} can be written compactly in matrix form as:

\begin{equation}
	\hat{\alpha}=\arg \min _\alpha\left\|(D \alpha-y) w_1\right\|_2^2+c\left\|w_2 \alpha\right\|_1.
\end{equation}

The optimization procedure based on ADMM is outlined in Algorithm \ref{alg:sar_despeckling}.

\begin{table}[t]
	\caption{Detailed information of SAR datasets\label{tb1}}
	\scriptsize
	\centering
	\newcolumntype{C}{>{\centering\arraybackslash}X}
	
	\begin{tabularx}{0.8\linewidth}{>{\raggedright\arraybackslash}p{1.6cm}CCC}
		\toprule
		\textbf{Method} & \textbf{Band} & \textbf{Size} & \textbf{Resolution(m)} \\
		\midrule
		Sentinel-1 & C & 256$\times$256 & 5-25 \\
		Gaofen-3   & C & 512$\times$512 & 3-25 \\
		TerraSAR-X  & X & 512$\times$512 & 3 \\
		miniSAR & X & 1255$\times$819 & 0.1 \\
		\multirow{2}{*}{FARAD} & X & 1434$\times$1007 & 0.1 \\
		& Ka & 650$\times$1020 & 0.1 \\
		\bottomrule
	\end{tabularx}
\end{table}

\subsection{Adaptive Noise Estimation}
\label{sec:sigma}
After the Log-Yeo--Johnson transformation, the multiplicative gamma 
noise is mapped to an approximately additive Gaussian field 
whose variance remains spatially non-stationary: it depends on 
the local backscatter intensity and on how well the gamma-to-Gaussian 
mapping holds in each region. A single global noise level therefore 
cannot describe the transformed field, which directly motivates the 
per-group weighting matrix $w_1$. We estimate the noise in two stages.

\noindent \textbf{Global initialization: } 
On the transformed image $\tilde{y}$ we obtain an image-level estimate 
$\hat{\sigma}_0$ by applying the MAD estimator to the diagonal detail 
wavelet sub-band $c_D$ (HH band), which is dominated by high-frequency 
noise:
\begin{equation}
	\hat{\sigma}_0 = \kappa \cdot \mathrm{median}\big(|c_D|\big),
\end{equation}
where $\kappa$ is the standard MAD-to-Gaussian calibration constant. 
$\hat{\sigma}_0$ initializes the optimization and provides a floor 
$\sigma_{\min}=\eta\,\hat{\sigma}_0$ ($\eta=0.1$) that prevents 
ill-conditioning of $w_1$.

\begin{figure}[!t]
	\centering
	
	\subfloat[]{
		\includegraphics[width=\linewidth]{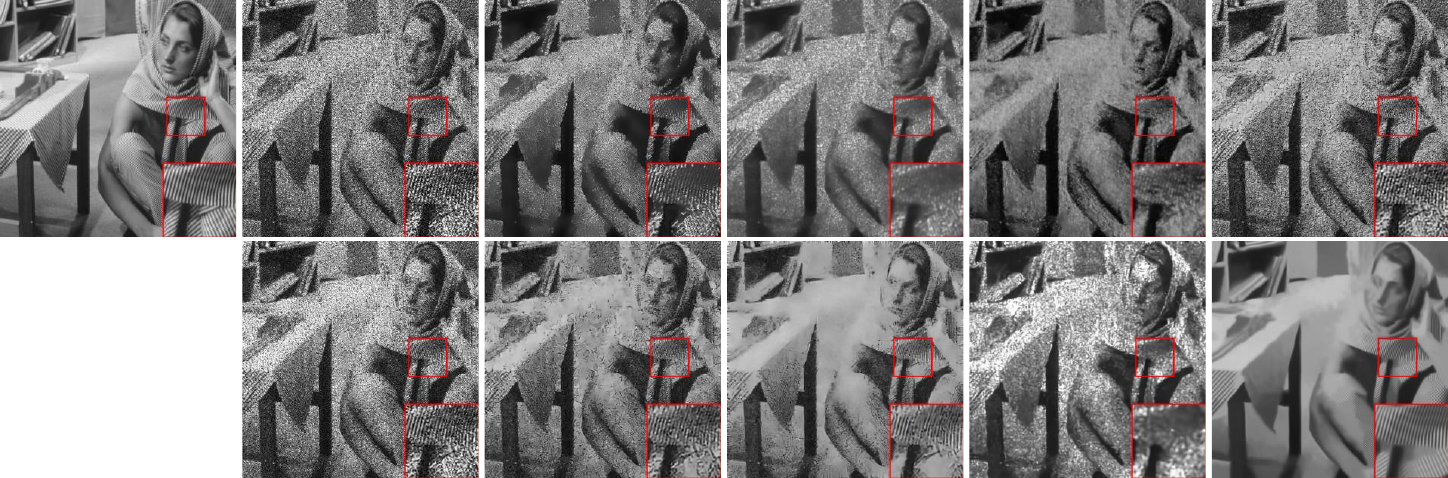}
		\label{fig_2}
	}
	
	\vspace{-0.7em}
	
	\subfloat[]{
		\includegraphics[width=\linewidth]{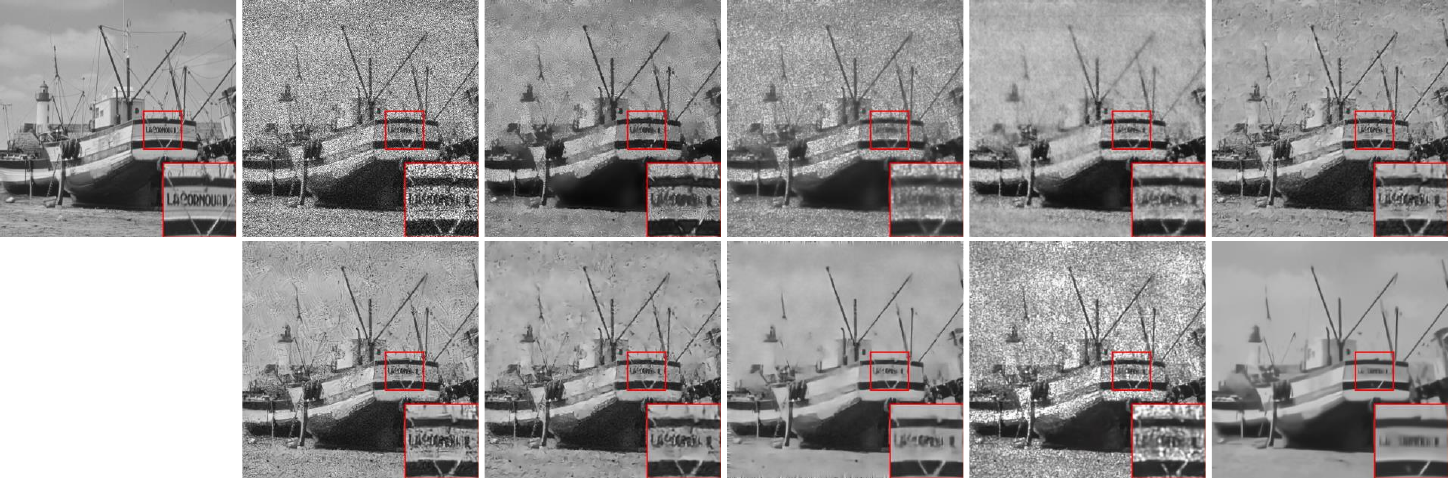}
		\label{fig_3}
	}
	
	\vspace{-0.7em}
	
	\subfloat[]{
		\includegraphics[width=\linewidth]{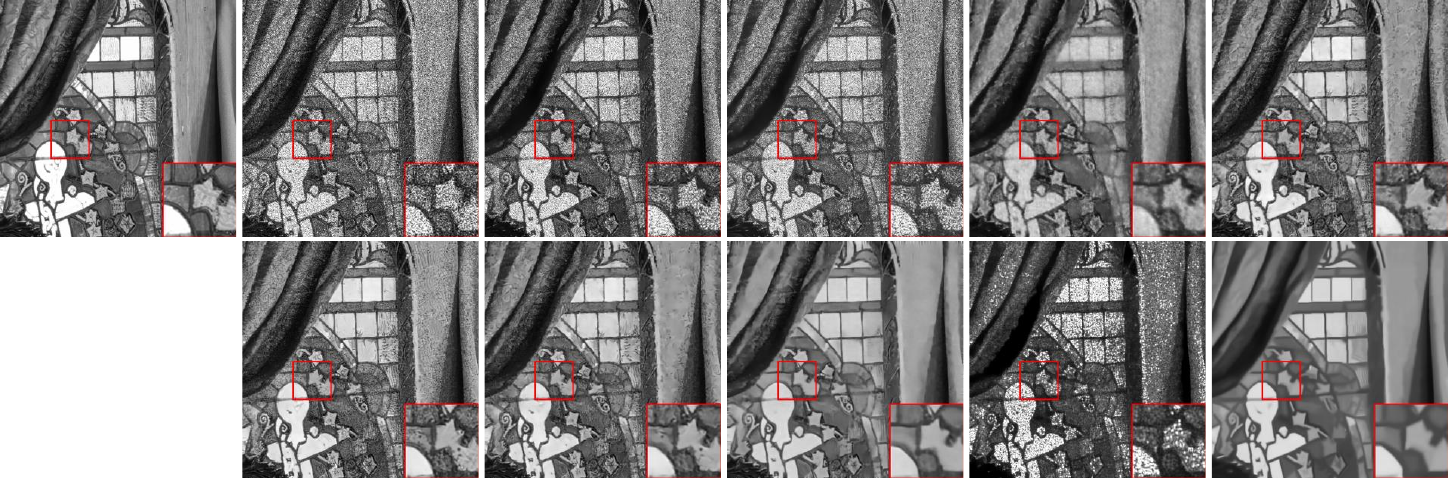}
		\label{fig_4}
	}
	
	\caption{Despeckling results on synthetic images. (a) Set12 (1-Look), (b) Set12 (2-Look), (c) McMaster (4-Look). Columns from right to left: the first row: ground-truth, noisy, ANLM, DnCNN, SAR2SAR, SARCAM; the second row: AGSDNet, SIFSDNet, MONet, RDDPM, and our proposed method.}
	\label{fig:multi_gamma}
\end{figure}

\begin{figure}[t]
	\centering
	\includegraphics[width=\linewidth]{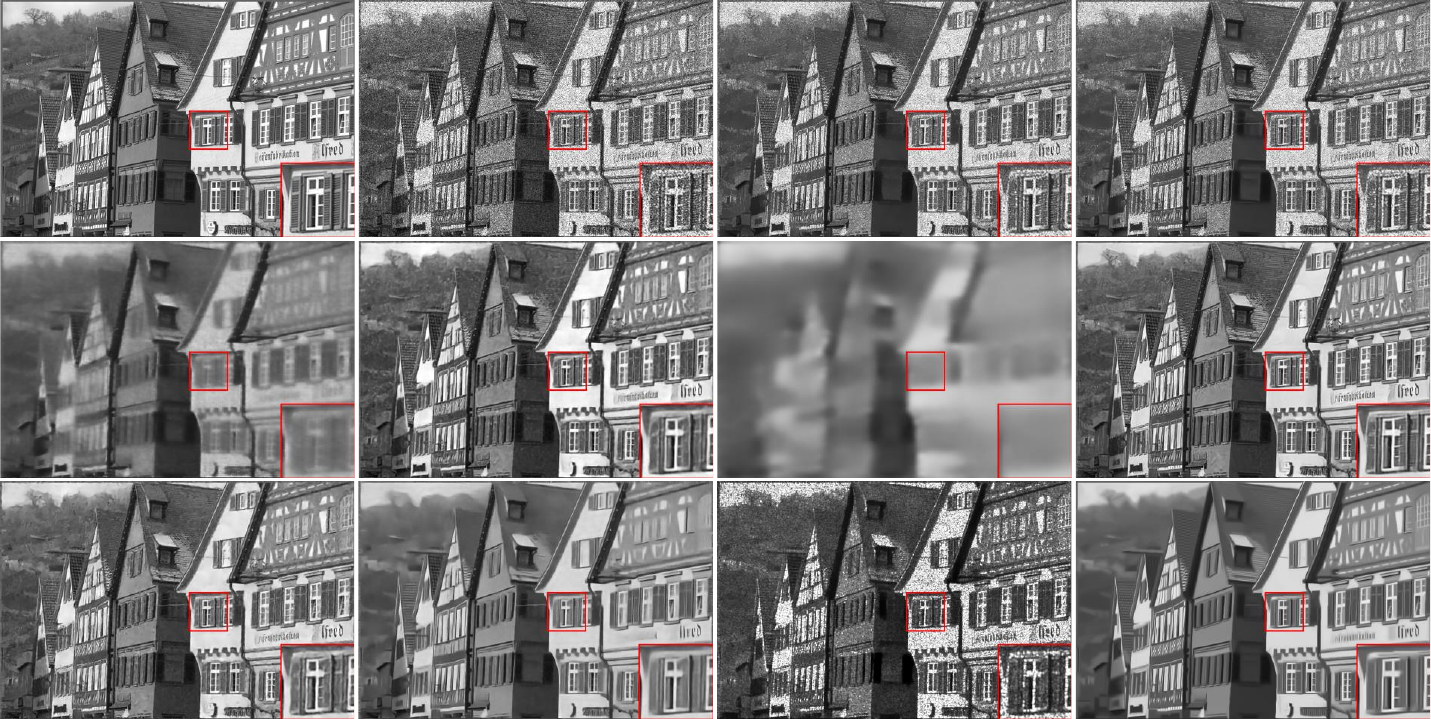}
	\caption{Despeckling results on Kodak24 synthetic images, column from right to left: the first row: ground-truth, noisy, ANLM, DnCNN; the second row: SAR2SAR, SARCAM, SARtrans, AGSDNet; the third row: , SIFSDNet, MONet, RDDPM, and our proposed method.}
	\label{fig_5}
\end{figure}

\noindent \textbf{Per-group residual refinement:} 
For a non-local group $Y=[y_1,\dots,y_K]\in\mathbb{R}^{p^2\times K}$ 
and the current sparse codes $\alpha^{(t)}$, let 
$r_k^{(t)}=y_k - D\alpha_k^{(t)}$ be the residual of the $k$-th patch 
at outer iteration $t$. To remain robust against structural leakage 
(e.g.\ edges that survive the sparse approximation), we re-apply the 
MAD estimator to $r_k^{(t)}$:
\begin{equation}
	\hat{\sigma}_k^{(t)} 
	= \max\!\left(\mathrm{MAD}\big(r_k^{(t)}\big),\; \sigma_{\min}\right), 
	\quad k=1,\dots,K,
	\label{eq:sigmak}
\end{equation}
and update $w_1^{(t)} = \mathrm{diag}(1/\hat{\sigma}_1^{(t)},\dots,
1/\hat{\sigma}_K^{(t)})$. Jointly refined with $\alpha$ in the outer 
loop, this closed-loop estimation progressively down-weights heavily 
speckled patches (relaxed fidelity, suppressing pseudo-structures) 
while preserving strong fidelity on clean patches, making $w_1$ 
genuinely region-aware.

  \begin{table}[H]
	\scriptsize
	\centering
	\caption{Quantitative Experiment on Set12 Synthetic Images\label{tb2}}
	\begin{tabularx}{\linewidth}{p{1.6cm}*{8}{>{\centering\arraybackslash}X}}
		\toprule[1pt]
		\multirow{2}{*}{} & \multicolumn{2}{c}{\textbf{1-Look}} & \multicolumn{2}{c}{\textbf{2-Look}} & \multicolumn{2}{c}{\textbf{4-Look}} & \multicolumn{2}{c}{\textbf{8-Look}}  \\
		& \multicolumn{1}{>{\hsize=1.5\hsize}X}{\textbf{PSNR}} & \multicolumn{1}{>{\hsize=1.5\hsize}X}{\textbf{SSIM}} & \multicolumn{1}{>{\hsize=1.5\hsize}X}{\textbf{PSNR}} &
		\multicolumn{1}{>{\hsize=1.5\hsize}X}{\textbf{SSIM}} & \multicolumn{1}{>{\hsize=1.5\hsize}X}{\textbf{PSNR}} & \multicolumn{1}{>{\hsize=1.5\hsize}X}{\textbf{SSIM}} & \multicolumn{1}{>{\hsize=1.5\hsize}X}{\textbf{PSNR}} &
		\multicolumn{1}{>{\hsize=1.5\hsize}X}{\textbf{SSIM}} \\
		\midrule
		ANLM\cite{xiao2020asymptotic} & 14.11 & 27.86 & 17.97 & 40.04 & 23.04 & 60.08 & 22.94 & 66.98 \\
		DIP \cite{ulyanov2018deep} & 17.08 & 48.69 & 18.18 & 56.81 & 19.04 & 61.86 & 20.83 & 68.97 \\
		DnCNN \cite{zhang2017beyond} & 18.36 & 39.91 & 20.72 & 46.02 & 22.43 & 51.99 & 23.44 & 58.01 \\
		SARCAM\cite{9633208} & 16.25 & 27.15 & 22.56 & 57.74 & 24.29 & 68.34 & 23.41 & 70.61 \\
		SAR2SAR\cite{dalsasso2021sar2sar} & 17.54 & 45.06 & 21.78 & 55.39 & 22.08 & 59.74 & 22.22 & 61.72 \\
		SARtrans \cite{perera2022transformer} & 18.84 & 51.58 & 19.11 & 52.15 & 18.71 & 52.30 & 18.55 & 52.51 \\
		AGSDNet\cite{thakur2022agsdnet} & 13.46 & 18.71 & 21.29 & 48.10 & \textbf{26.14} & \underline{75.33} & \textbf{27.76} & \underline{81.23} \\
		SIFSDNet\cite{thakur2022sifsdnet} & 18.66 & 38.67 & \underline{23.11} & 61.95 & 25.14 & 72.01 & 25.82 & 76.87 \\
		MONet\cite{vitale2023sar} & \underline{21.05} & \underline{54.49} & 22.77 & \underline{65.74} & 23.50 & 66.87 & 24.02 & 67.26 \\
		RDDPM\cite{hu2024sar} & 16.98 & 32.22 & 18.50 & 38.51 & 19.93 & 44.26 & 21.03 & 49.87 \\
		CL-SAR \cite{fang2024contrastive} & 15.52 & 29.07 & 16.61 & 55.30 & 17.97 & 71.16 & 19.11 & 77.11 \\
		S3DIP \cite{albisani2025self} & 16.38 & 50.98 & 17.30 & 56.06 & 17.00 & 62.57 & 16.26 & 64.35 \\
		Ours & \textbf{23.31} & \textbf{66.08} & \textbf{24.83} & \textbf{71.78} & \underline{25.98} & \textbf{77.13} & \underline{26.92} & \textbf{82.04} \\
		\bottomrule[1pt]
	\end{tabularx}
\end{table}

\vspace{-0.7em}

\begin{table}[H]
	\scriptsize
	\centering
	\caption{Quantitative Experiment on McMaster Synthetic Images\label{tb3}}
	\begin{tabularx}{\linewidth}{p{1.6cm}*{8}{>{\centering\arraybackslash}X}}
		\toprule[1pt]
		\multirow{2}{*}{} & \multicolumn{2}{c}{\textbf{1-Look}} & \multicolumn{2}{c}{\textbf{2-Look}} & \multicolumn{2}{c}{\textbf{4-Look}} & \multicolumn{2}{c}{\textbf{8-Look}}  \\
		& \multicolumn{1}{>{\hsize=1.5\hsize}X}{\textbf{PSNR}} & \multicolumn{1}{>{\hsize=1.5\hsize}X}{\textbf{SSIM}} & \multicolumn{1}{>{\hsize=1.5\hsize}X}{\textbf{PSNR}} &
		\multicolumn{1}{>{\hsize=1.5\hsize}X}{\textbf{SSIM}} & \multicolumn{1}{>{\hsize=1.5\hsize}X}{\textbf{PSNR}} & \multicolumn{1}{>{\hsize=1.5\hsize}X}{\textbf{SSIM}} & \multicolumn{1}{>{\hsize=1.5\hsize}X}{\textbf{PSNR}} &
		\multicolumn{1}{>{\hsize=1.5\hsize}X}{\textbf{SSIM}} \\
		\midrule
		ANLM\cite{xiao2020asymptotic} & 11.72 & 22.21 & 13.91 & 30.65 & 16.37 & 39.98 & 19.37 & 50.49 \\
		DIP \cite{ulyanov2018deep} & 16.09 & 52.75 & 18.70 & 61.90 & 20.29 & 67.68 & 22.46 & 72.67 \\
		DnCNN \cite{zhang2017beyond} & 11.83 & 22.86 & 13.86 & 31.98 & 16.56 & 43.12 & 20.08 & 54.77 \\
		SARCAM\cite{9633208} & 14.33 & 26.44 & 17.78 & 39.64 & 24.08 & 64.60 & 27.43 & 79.20 \\
		SAR2SAR\cite{dalsasso2021sar2sar} & 13.16 & 35.75 & 20.59 & 54.99 & 22.46 & 64.93 & 23.09 & 68.59 \\
		SARtrans \cite{perera2022transformer} & \underline{18.40} & 52.42 & 19.56 & 53.22 & 19.85 & 53.63 & 19.86 & 53.83 \\
		AGSDNet\cite{thakur2022agsdnet} & 12.18 & 18.11 & 15.14 & 27.49 & 22.66 & 55.87 & \textbf{28.49} & 81.67 \\
		SIFSDNet\cite{thakur2022sifsdnet} & 16.01 & 30.62 & 20.04 & 48.33 & \underline{24.90} & 69.54 & \underline{28.33} & \textbf{82.02} \\
		MONet\cite{vitale2023sar} & 15.15 & 26.57 & \underline{21.48} & 60.08 & 24.44 & \textbf{77.36} & 26.06 & 78.75 \\
		RDDPM\cite{hu2024sar} & 13.31 & 25.71 & 14.88 & 30.01 & 16.42 & 33.76 & 17.71 & 36.77 \\
		CL-SAR \cite{fang2024contrastive} & 15.77 & 36.12 & 17.11 & 60.60 & 18.95 & \underline{76.29} & 20.37 & \underline{81.75} \\
		S3DIP \cite{albisani2025self} & 14.96 & \underline{55.98} & 15.14 & \underline{64.40} & 15.15 & 64.85 & 15.30 & 62.62 \\
		Ours & \textbf{20.75} & \textbf{62.31} & \textbf{23.10} & \textbf{66.59} & \textbf{24.92} & 70.64 & 26.54 & 74.76 \\
		\bottomrule[1pt]
	\end{tabularx}
\end{table}

\vspace{-0.7em}

\begin{table}[H]
	\scriptsize
	\centering
	\caption{Quantitative Experiment on Kodak24 Synthetic Images\label{tb4}}
	\begin{tabularx}{\linewidth}{p{1.6cm}*{8}{>{\centering\arraybackslash}X}}
		\toprule[1pt]
		\multirow{2}{*}{} & \multicolumn{2}{c}{\textbf{1-Look}} & \multicolumn{2}{c}{\textbf{2-Look}} & \multicolumn{2}{c}{\textbf{4-Look}} & \multicolumn{2}{c}{\textbf{8-Look}}  \\
		& \multicolumn{1}{>{\hsize=1.5\hsize}X}{\textbf{PSNR}} & \multicolumn{1}{>{\hsize=1.5\hsize}X}{\textbf{SSIM}} & \multicolumn{1}{>{\hsize=1.5\hsize}X}{\textbf{PSNR}} &
		\multicolumn{1}{>{\hsize=1.5\hsize}X}{\textbf{SSIM}} & \multicolumn{1}{>{\hsize=1.5\hsize}X}{\textbf{PSNR}} & \multicolumn{1}{>{\hsize=1.5\hsize}X}{\textbf{SSIM}} & \multicolumn{1}{>{\hsize=1.5\hsize}X}{\textbf{PSNR}} &
		\multicolumn{1}{>{\hsize=1.5\hsize}X}{\textbf{SSIM}} \\
		\midrule
		ANLM\cite{xiao2020asymptotic} & 11.29 & 10.03 & 13.49 & 16.32 & 16.13 & 26.33 & 19.27 & 40.46 \\
		DIP \cite{ulyanov2018deep} & \underline{19.01} & 50.47 & 20.51 & 55.32 & 22.02 & 59.86 & 23.37 & 64.39 \\
		DnCNN \cite{zhang2017beyond} & 11.41 & 10.30 & 13.35 & 16.55 & 16.04 & 27.60 & 19.62 & 43.76 \\
		SARCAM\cite{9633208} & 13.85 & 15.47 & 16.97 & 26.39 & 22.64 & 51.64 & 25.38 & 69.12 \\
		SAR2SAR\cite{dalsasso2021sar2sar} & 16.95 & 49.35 & 20.10 & 55.39 & 20.76 & 57.76 & 20.95 & 59.13 \\
		SARtrans \cite{perera2022transformer} & 17.95 & 48.42 & 19.14 & 48.91 & 19.63 & 49.13 & 19.77 & 49.19 \\
		AGSDNet\cite{thakur2022agsdnet} & 11.80 & 10.60 & 14.66 & 17.55 & 21.29 & 43.05 & \textbf{26.30} & 72.83 \\
		SIFSDNet\cite{thakur2022sifsdnet} & 15.55 & 19.81 & 19.19 & 36.09 & \underline{23.48} & 58.20 & \textbf{26.30} & \underline{73.14} \\
		MONet\cite{vitale2023sar} & 14.71 & 17.44 & \underline{20.67} & 47.89 & 23.45 & \underline{68.00} & 24.59 & 67.78 \\
		RDDPM\cite{hu2024sar} & 13.03 & 21.41 & 14.60 & 26.34 & 16.08 & 30.85 & 17.40 & 34.36 \\
		CL-SAR \cite{fang2024contrastive} & 15.91 & 23.78 & 18.37 & 50.32 & 20.36 & \textbf{68.74} & 22.36 & \textbf{75.09} \\
		S3DIP \cite{albisani2025self} & 17.47 & \underline{51.96} & 16.50 & \underline{58.01} & 15.91 & 58.02 & 16.02 & 56.57 \\
		Ours & \textbf{20.11} & \textbf{55.86} & \textbf{22.20} & \textbf{59.31} & \textbf{23.73} & 62.67 & \underline{25.52} & 66.31 \\
		\bottomrule[1pt]
	\end{tabularx}
\end{table}

\section{Results and Analysis}

\subsection{Experimental Setup}

\noindent \textbf{Dataset:} To evaluate the performance of different despeckling methods, we carried out comprehensive qualitative and quantitative experiments on eight datasets, covering both synthetic noisy images and real SAR data. In synthetic experiments, we selected three commonly used benchmark datasets: Set12, McMaster, and Kodak24. These datasets contain optical images, which were corrupted with gamma noise at different equivalent number of looks. After adding synthetic noise, the image histograms were adjusted to approximate the mean of original images, better simulating SAR image characteristics.

For real SAR despeckling evaluation, we tested on data from various sources, bands, and resolutions to demonstrate the adaptability of our method in different scenarios. In particular, we used Sentinel-1 \cite{torres2012gmes}, Gaofen-3 \cite{zhao2021china}, TerraSAR-X, miniSAR, and FARAD data. Table \ref{tb1} summarizes the imaging characteristics of SAR images from these sources.

\noindent \textbf{Evaluation Metrics:} In synthetic experiments, ground-truth noise-free images are available, allowing us to use Peak Signal-to-Noise Ratio (PSNR) and Structural Similarity Index Measure (SSIM) as evaluation metrics. For real-world SAR images, we adopted Equivalent Number of Looks (ENL), Edge Preservation Index (EPI), Edge Preservation Degree Based on the Ratio of the Average (EPD-ROA), Structural Quality Index (SQI), and mean intensity as evaluation criteria.

\noindent \textbf{Implementation Specifics}: For Log-Yeo-Johnson transformation, we employed an exhaustive search to determine the optimal parameter $\lambda$, ensuring that the transformed data distribution closely approximates Gaussian distribution. Our proposed method jointly processes similar image patches within a region, using a patch size of 16 $\times$ 16 and a stacking count of 10. We set the regularization coefficient $c$ to 1.5 to enforce sparsity in sparse coding process.

\noindent \textbf{Baseline Method}: To ensure a fair comparison, we evaluated nine different SAR despeckling methods, all using the official parameter settings and pre-trained models provided by the authors. The compared methods include ANLM \cite{xiao2020asymptotic}, DIP \cite{ulyanov2018deep}, DnCNN \cite{zhang2017beyond}, SAR2SAR \cite{dalsasso2021sar2sar}, SARCAM \cite{9633208}, SARtrans \cite{perera2022transformer}, AGSDNet \cite{thakur2022agsdnet}, SIFSDNet \cite{thakur2022sifsdnet}, MONet \cite{vitale2021analysis}, CL-SAR \cite{fang2024contrastive}, S3DIP \cite{albisani2025self}, and RDDPM \cite{hu2024sar}.

\begin{table}[t]
	\caption{Quantitative Experiment on Sentinel-1 Samples (Mean = 34.39)\label{tb6}}
	\scriptsize
	\centering
	\newcolumntype{C}{>{\centering\arraybackslash}X}
	
	\begin{tabularx}{\linewidth}{>{\raggedright\arraybackslash}p{1.6cm}CCCCCC}
		\toprule
		\textbf{Method} & \textbf{ENL} & \textbf{EPI} & \textbf{EPD(H)} & \textbf{EPD(V)} & \textbf{SQI} & \textbf{Mean} \\
		\midrule
		ANLM \cite{xiao2020asymptotic}    & 1.623 & \underline{0.719} & 13.14 & 13.01 & \textbf{0.787} & 29.99 \\
		DIP \cite{ulyanov2018deep} & 2.766 & 0.117 & 11.15 & 10.68 & 0.601 & 31.89 \\
		DnCNN\cite{zhang2017beyond}   & 2.781 & 0.618 & 11.31 & 11.46 & 0.589 & 35.67 \\
		SARCAM\cite{9633208}  & 2.816 & 0.621 & 11.27 & 11.26 & 0.595 & \underline{34.75} \\
		SAR2SAR\cite{dalsasso2021sar2sar} & 2.868 & 0.651 & 6.61  & 6.34  & 0.645 & 27.46 \\
		AGSDNet\cite{thakur2022agsdnet} & 2.738 & 0.626 & \underline{13.89} & \underline{14.03} & 0.600 & 34.84 \\
		SIFSDNet\cite{thakur2022sifsdnet} & 3.019 & 0.599 & 12.67 & 12.68 & 0.567 & 35.70 \\
		MONet\cite{vitale2023sar}   & \textbf{4.548} & 0.497 & 7.66  & 7.51  & 0.455 & 37.71 \\
		RDDPM \cite{hu2024sar}  & 2.795 & 0.614 & 12.36 & 12.19 & 0.583 & 37.17 \\
		CL-SAR \cite{fang2024contrastive} & 2.802 & 0.226 & 13.45 & 13.30 & 0.619 & 37.90 \\
		S3DIP \cite{albisani2025self} & \underline{3.767} & 0.081 & 2.98  & 2.57  & 0.643 & 44.35 \\
		Ours    & 2.784 & \textbf{0.733} & \textbf{14.33} & \textbf{14.32} & \underline{0.695} & \textbf{34.63} \\
		\bottomrule
	\end{tabularx}
\end{table}

\subsection{Synthetic Experiment}


We conducted experiments on synthetic noisy images using three commonly adopted benchmark datasets: Set12, McMaster, and Kodak24. 

For the McMaster and Kodak24 datasets, we converted the images to grayscale to simulate the single-channel nature of SAR images. Gamma noise with ENL values of 1, 2, 4, and 8 was added for testing, in order to evaluate the adaptability and effectiveness of various methods under different noise levels. Figures \ref{fig_2} and \ref{fig_3} present a visual comparison of despeckling results under high-intensity noise conditions with ENL = 1 and 2. It can be observed that the proposed method achieves the best overall performance. Benefiting from the sparsity representation property, the model effectively restores structural information and fine details (e.g., the striped patterns on the clothes in Figure \ref{fig_2}), while demonstrating superior noise suppression and smoother despeckled results compared to other methods. Other despeckling methods  and deep learning-based approaches tend to be limited by parameter selection and overfitting, and thus fail to effectively remove high-intensity gamma noise.

%

\begin{figure*}[t]
	\centering
	\includegraphics[width=0.86\textwidth]{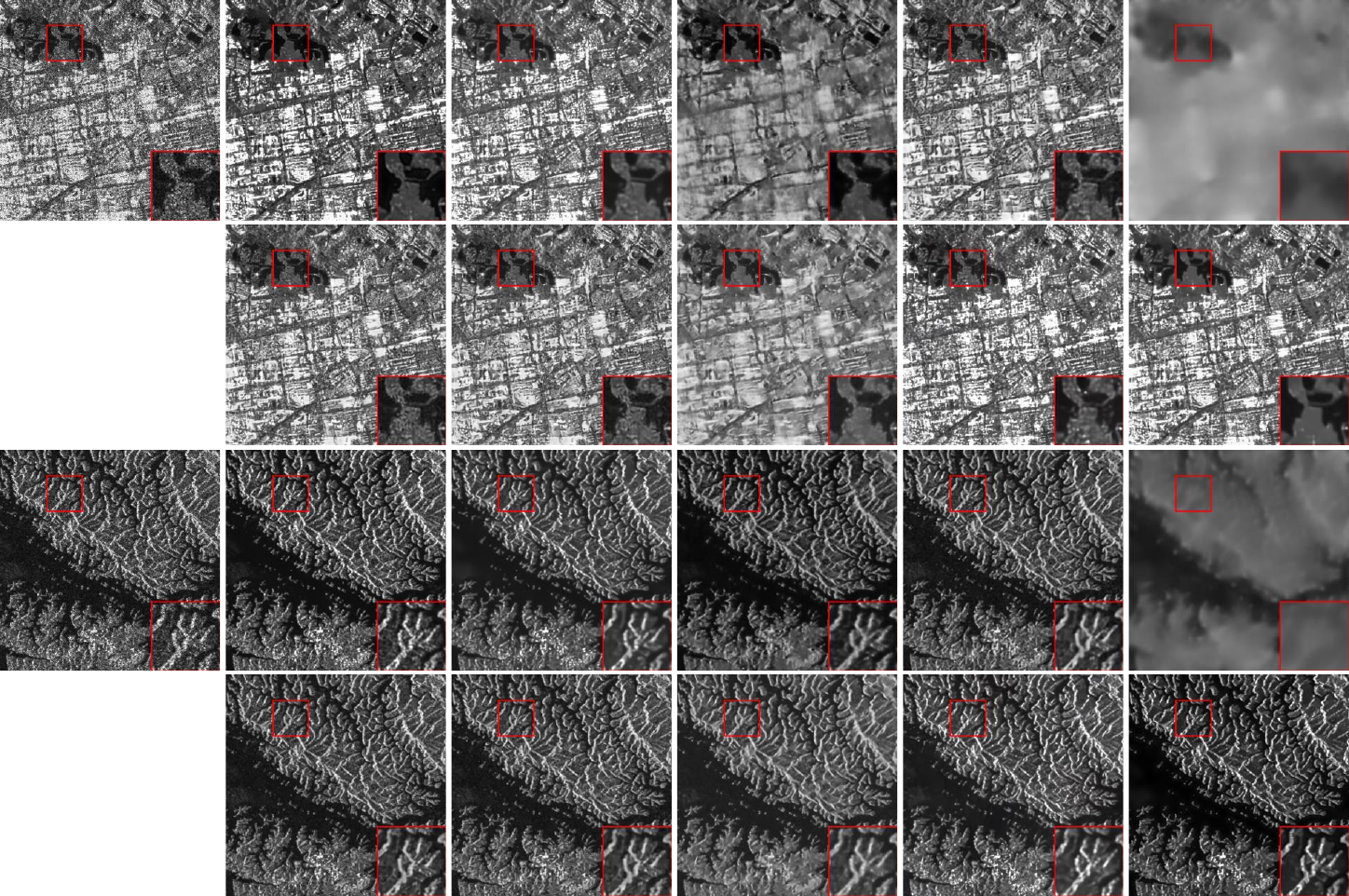}
	\caption{SAR image despeckling on Sentinel-1, Columns from right to left: the first row: noisy, ANLM, DnCNN, SAR2SAR, SARCAM, SARtrans; the second row: AGSDNet, SIFSDNet, MONet, RDDPM, and our proposed method.}
	\label{fig_6}
\end{figure*}

%

\begin{table}[t]
	\caption{Quantitative Experiment on Gaofen-3 Samples (Mean = 20.10)\label{tb7}}
	\scriptsize
	\centering
	\newcolumntype{C}{>{\centering\arraybackslash}X}
	
	\begin{tabularx}{\linewidth}{>{\raggedright\arraybackslash}p{1.6cm}CCCCCC}
		\toprule
		\textbf{Method} & \textbf{ENL} & \textbf{EPI} & \textbf{EPD(H)} & \textbf{EPD(V)} & \textbf{SQI} & \textbf{Mean} \\
		\midrule
		ANLM\cite{xiao2020asymptotic}    & 0.930 & 0.609 & 34.65 & 33.25 & 0.646 & 20.87 \\
		DIP \cite{ulyanov2018deep} & 0.670 & 0.081 & 22.12 & 20.94 & 0.445 & 18.37 \\
		DnCNN\cite{zhang2017beyond}   & 1.177 & 0.689 & 28.36 & 27.73 & 0.766 & 20.23 \\
		SARCAM\cite{9633208}  & 1.231 & 0.603 & 17.35 & 17.47 & 0.784 & \underline{20.01} \\
		SAR2SAR\cite{dalsasso2021sar2sar} & 1.103 & 0.474 & 10.05 & 10.14 & 0.747 & 16.67 \\
		AGSDNet\cite{thakur2022agsdnet} & 1.137 & \textbf{0.847} & \underline{43.71} & \underline{44.23} & 0.743 & 19.89 \\
		SIFSDNet\cite{thakur2022sifsdnet} & 1.290 & 0.721 & 25.02 & 24.99 & \underline{0.812} & 20.86 \\
		MONet\cite{vitale2023sar}   & \textbf{1.577} & 0.638 & 11.42 & 10.95 & \textbf{0.823} & 21.96 \\
		RDDPM \cite{hu2024sar}  & 0.788 & 0.661 & 32.07 & 29.61 & 0.657 & 18.82 \\
		CL-SAR \cite{fang2024contrastive} & 1.057 & 0.200 & 43.47 & 44.39 & 0.498 & 20.37 \\
		S3DIP \cite{albisani2025self} & 0.909 & 0.065 & 5.97  & 5.90  & 0.481 & 22.24 \\
		Ours    & \underline{1.297} & \underline{0.781} & \textbf{48.85} & \textbf{47.13} & 0.768 & \textbf{20.10} \\
		\bottomrule
	\end{tabularx}
\end{table}

\begin{table}[t]
	\caption{Quantitative Experiment on TerraSAR-X (Mean = 30.04)\label{tb8}}
	\scriptsize
	\centering
	\newcolumntype{C}{>{\centering\arraybackslash}X}
	
	\begin{tabularx}{\linewidth}{>{\raggedright\arraybackslash}p{1.6cm}CCCCCC}
		\toprule
		\textbf{Method} & \textbf{ENL} & \textbf{EPI} & \textbf{EPD(H)} & \textbf{EPD(V)} & \textbf{SQI} & \textbf{Mean} \\
		\midrule
		ANLM\cite{xiao2020asymptotic}    & 1.413 & 0.553 & 45.01 & 40.85 & 0.676 & 29.13 \\
		DIP \cite{ulyanov2018deep} & 4.197 & 0.063 & 9.53  & 10.10 & 0.604 & 33.42 \\
		DnCNN\cite{zhang2017beyond}   & 1.546 & 0.550 & 41.22 & 38.55 & 0.732 & 30.09 \\
		SARCAM\cite{9633208}  & 1.696 & 0.548 & 21.22 & 21.76 & 0.787 & 30.85 \\
		SAR2SAR\cite{dalsasso2021sar2sar} & \underline{2.236} & 0.276 & 10.90 & 10.80 & \underline{0.867} & 22.85 \\
		AGSDNet\cite{thakur2022agsdnet} & 1.463 & \textbf{0.697} & \underline{46.88} & \underline{46.28} & 0.697 & \underline{30.60} \\
		SIFSDNet\cite{thakur2022sifsdnet} & 1.716 & 0.614 & 28.43 & 27.76 & 0.796 & 31.48 \\
		MONet\cite{vitale2023sar}   & \textbf{2.268} & 0.457 & 13.69 & 13.46 & \textbf{0.885} & 32.73 \\
		RDDPM \cite{hu2024sar}  & 1.194 & 0.495 & 33.97 & 31.42 & 0.572 & 28.60 \\
		CL-SAR \cite{fang2024contrastive} & 1.494 & 0.266 & 45.40 & 45.26 & 0.544 & 32.53 \\
		S3DIP \cite{albisani2025self} & 2.195 & 0.088 & 4.52  & 4.69  & 0.593 & 35.69 \\
		Ours    & 1.804 & \underline{0.682} & \textbf{54.44} & \textbf{50.84} & 0.784 & \textbf{30.05} \\
		\bottomrule
	\end{tabularx}
\end{table}

\begin{table}[t]
	\caption{Quantitative Experiment on miniSAR Samples (Mean = 13.32)\label{tb9}}
	\scriptsize
	\centering
	\newcolumntype{C}{>{\centering\arraybackslash}X}
	
	\begin{tabularx}{\linewidth}{>{\raggedright\arraybackslash}p{1.6cm}CCCCCC}
		\toprule
		\textbf{Method} & \textbf{ENL} & \textbf{EPI} & \textbf{EPD(H)} & \textbf{EPD(V)} & \textbf{SQI} & \textbf{Mean} \\
		\midrule
		ANLM\cite{xiao2020asymptotic}    & 4.438 & 0.172 & 3.52 & 4.20 & 0.519 & 9.82 \\
		DIP \cite{ulyanov2018deep} & 3.817 & 0.028 & 2.27 & 2.80 & 0.655 & 29.24 \\
		DnCNN\cite{zhang2017beyond}   & 8.538 & 0.141 & 2.16 & 2.66 & 0.726 & 13.52 \\
		SARCAM\cite{9633208}  & \textbf{9.204} & 0.023 & 3.03 & 3.21 & \underline{0.755} & 12.00 \\
		SAR2SAR\cite{dalsasso2021sar2sar} & 5.872 & 0.023 & 2.37 & 2.36 & 0.592 & 12.97 \\
		AGSDNet\cite{thakur2022agsdnet} & 6.013 & \textbf{0.638} & \underline{15.24} & \textbf{16.24} & 0.610 & \underline{13.24} \\
		SIFSDNet\cite{thakur2022sifsdnet} & 8.680 & 0.074 & 3.89 & 3.38 & 0.733 & 14.04 \\
		MONet\cite{vitale2023sar}   & 8.449 & 0.132 & 2.53 & 2.85 & 0.724 & 14.30 \\
		RDDPM \cite{hu2024sar}  & 2.385 & 0.545 & 13.63 & 14.47 & 0.382 & 11.65 \\
		CL-SAR \cite{fang2024contrastive} & 7.478 & 0.016 & 4.97  & 5.31  & 0.730 & 15.64 \\
		S3DIP \cite{albisani2025self} & 3.744 & 0.006 & 0.78  & 0.81  & 0.656 & 42.40 \\
		Ours    & \underline{9.008} & \underline{0.593} & \textbf{15.41} & \underline{15.01} & \textbf{0.804} & \textbf{13.31} \\
		\bottomrule
	\end{tabularx}
\end{table}

Figure \ref{fig_4} presents the despeckling results on McMaster dataset under moderate noise conditions (ENL = 4). The proposed method, along with MONet, effectively suppresses speckle noise and mitigates edge artifacts, delivering the best visual performance. In contrast, other methods still exhibit noticeable residual noise. Figure \ref{fig_5} shows the results under a lower noise level (ENL = 8). In this scenario, most methods achieve visually satisfactory results, while the proposed method maintains a good balance between noise suppression and detail preservation. Although our method introduces slight blurring, it still effectively removes noise and preserves the detailed features.



To avoid the impact of parameter tuning on model adaptability, we set the parameter $c$ to 1.5 for all synthetic image experiments. In subsequent parameter sensitivity studies, we demonstrate that adjusting $c$ also allows the model to be fine-tuned to balance denoising performance and detail preservation.

Tables \ref{tb2}, \ref{tb3}, and \ref{tb4} present the quantitative results on three datasets. Consistent with the qualitative experiments, our method significantly outperforms other methods under high-intensity noise conditions and also achieves satisfactory performance in low-intensity noise scenarios.

\begin{figure}[!t]
	\centering
	
	\subfloat[]{
		\includegraphics[width=\linewidth]{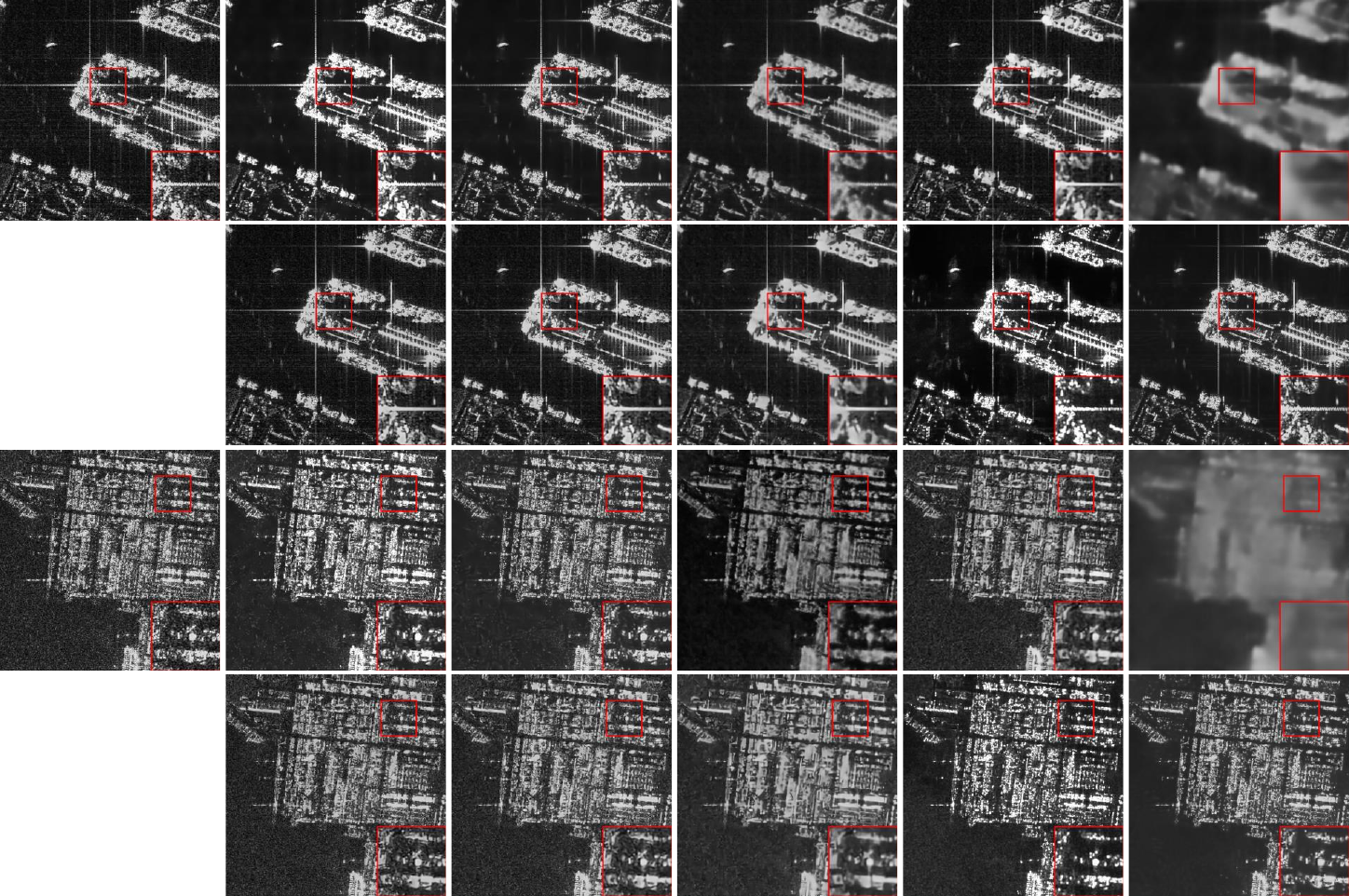}
		\label{fig_7}
	}
	
	\vspace{-0.7em}
	
	\subfloat[]{
		\includegraphics[width=\linewidth]{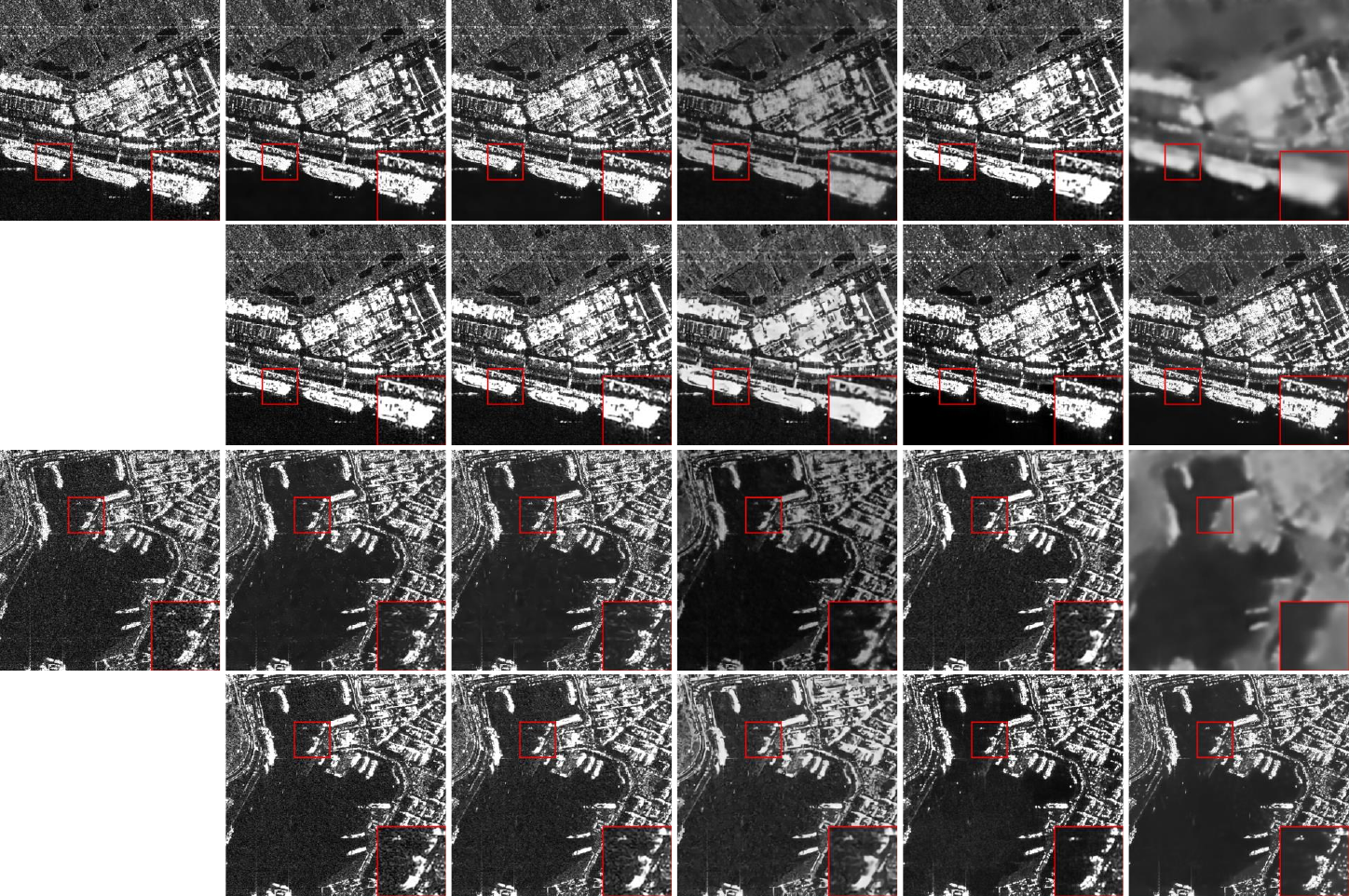}
		\label{fig_8}
	}
	
	\caption{SAR image despeckling on (a) Gaofen-3 and (b) TerraSAR-X, Columns from right to left: the first row: noisy, ANLM, DnCNN, SAR2SAR, SARCAM, SARtrans; the second row: AGSDNet, SIFSDNet, MONet, RDDPM, and our proposed method.}
\end{figure}

\subsection{Real Experiment}

\begin{figure*}[!t]
	\centering
	
	\subfloat[]{
		\includegraphics[width=0.86\textwidth]{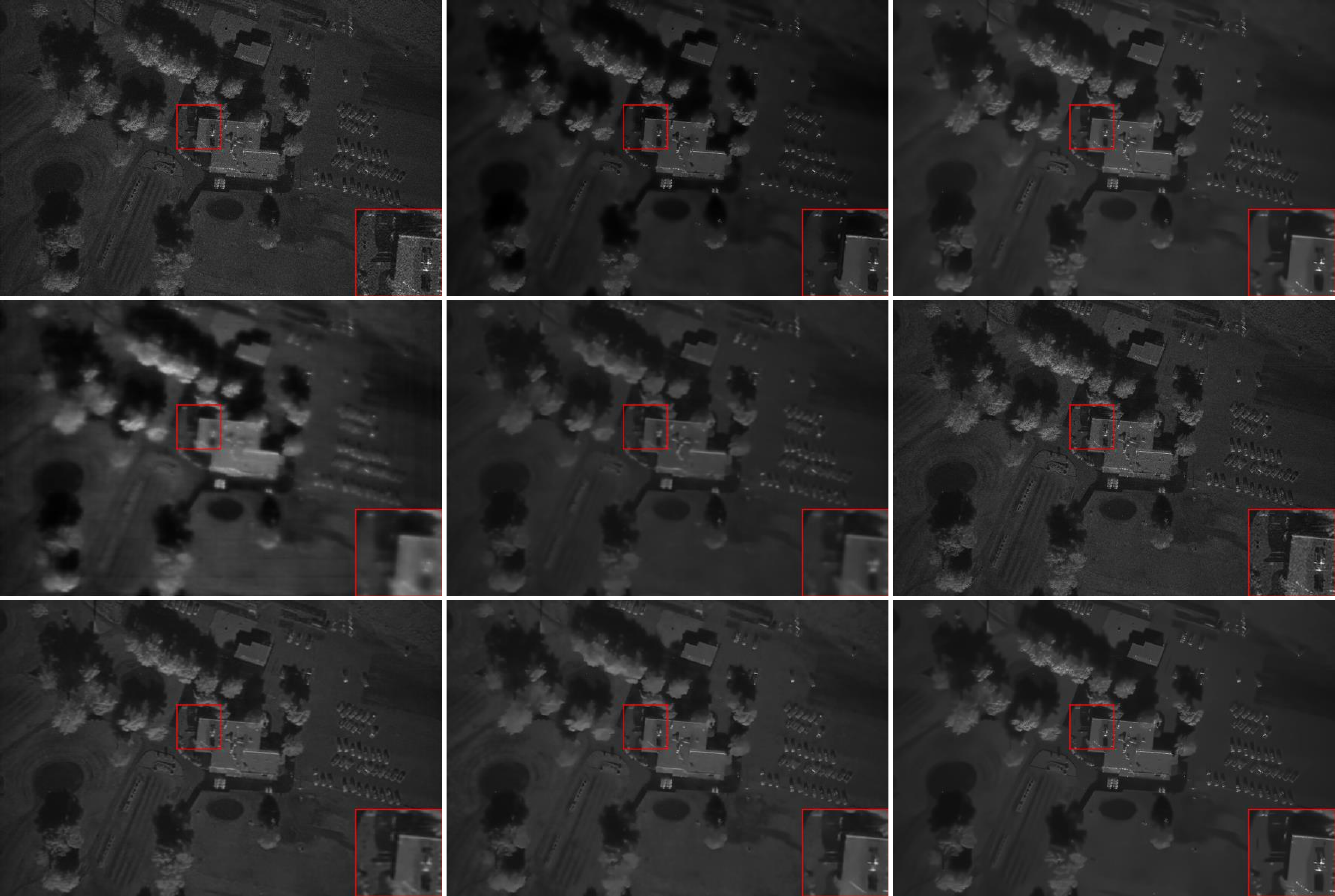}
		\label{fig_9}
	}
	
	\vspace{-0.7em}
	
	\subfloat[]{
		\includegraphics[width=0.86\textwidth]{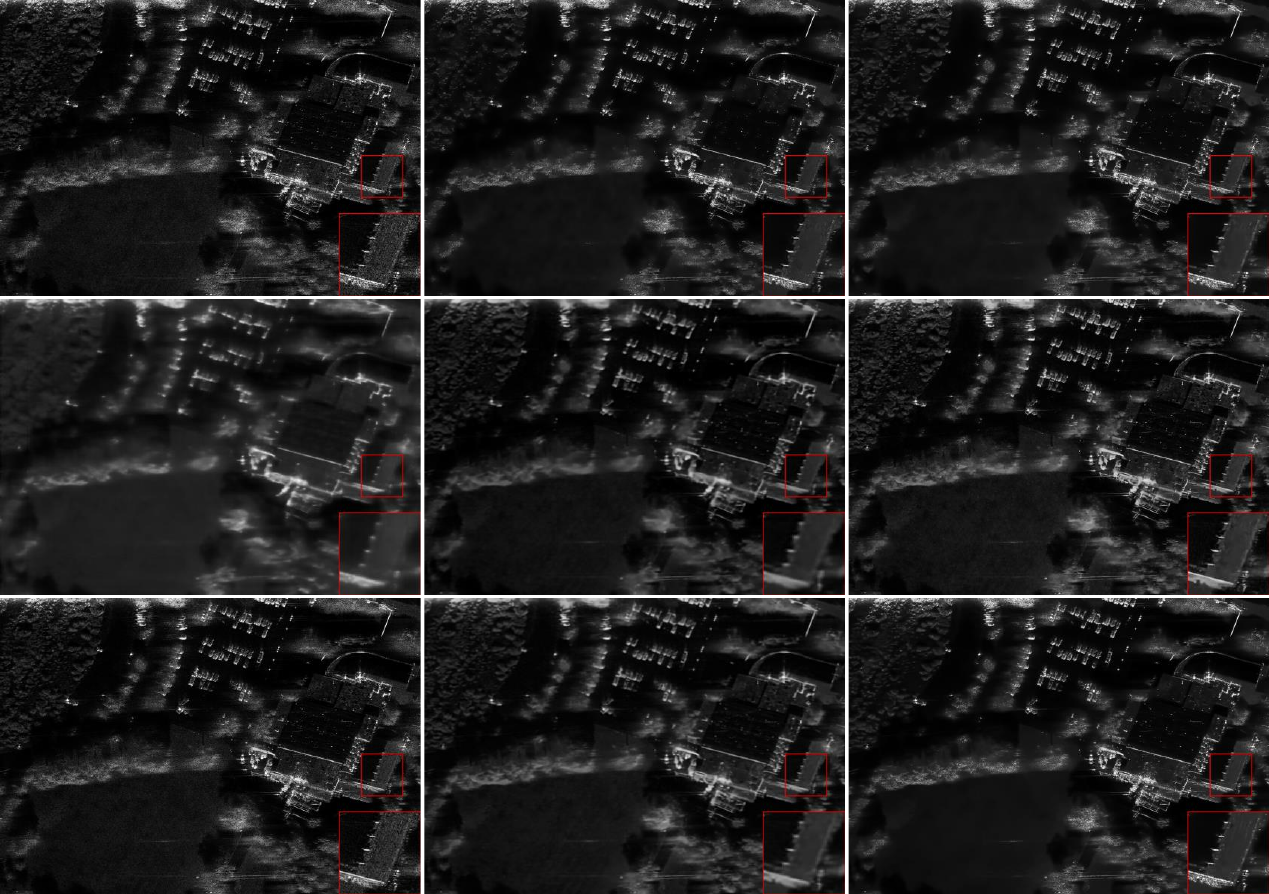}
		\label{fig_10}
	}
	
	\caption{SAR image despeckling on (a) miniSAR and (b) FARAD\_X, Columns from right to left: the first row: noisy, ANLM, DnCNN; the second row: SAR2SAR, SARCAM, AGSDNet; the third row: SIFSDNet, MONet, and our proposed method.}
\end{figure*}

\begin{figure*}[t]
	\centering
	\includegraphics[width=0.85\textwidth]{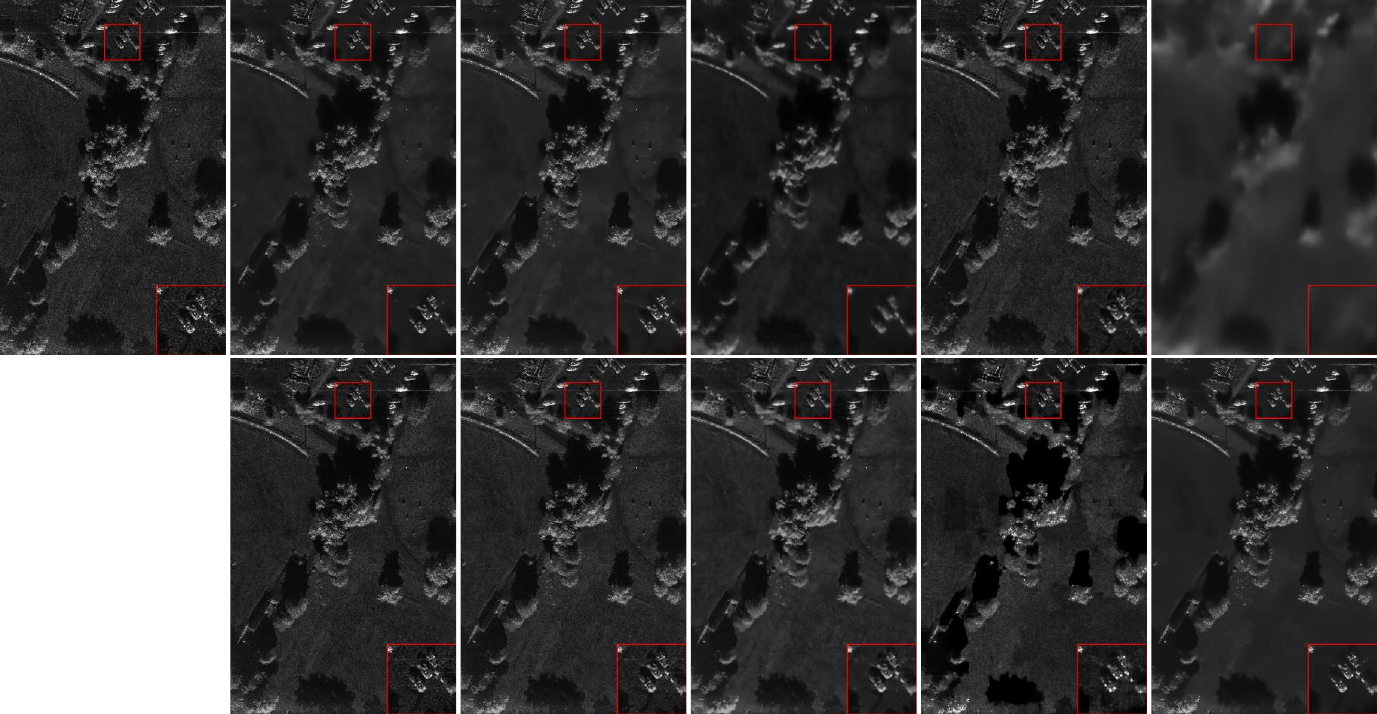}
	\caption{SAR image despeckling on FARAD\_Ka, Columns from right to left: the first row: noisy, ANLM, DnCNN, SAR2SAR, SARCAM, SARtrans; the second row: AGSDNet, SIFSDNet, MONet, RDDPM, and our proposed method.}
	\label{fig_11}
\end{figure*}

\begin{figure}[t]
	\centering
	\includegraphics[width=\linewidth]{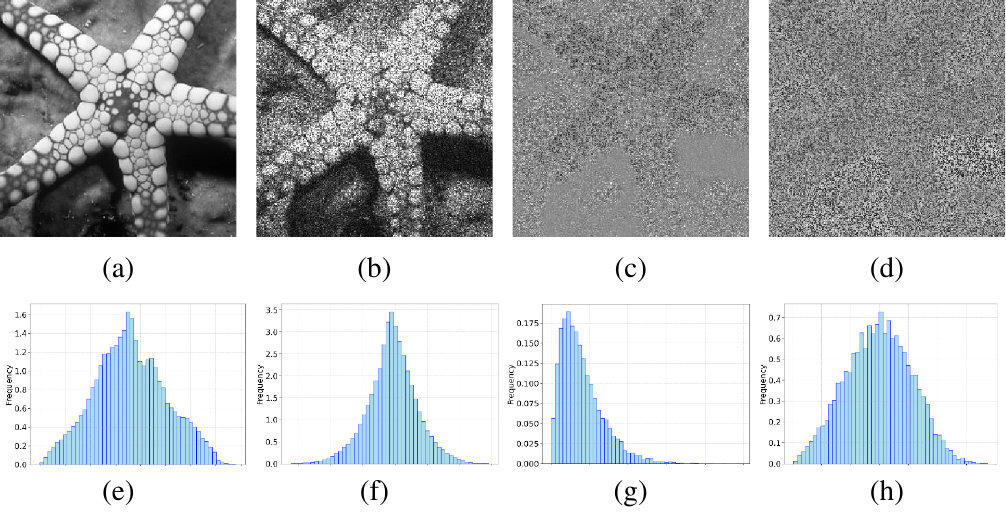}
	\caption{Images and Noise distributions before and after Log-Yeo-Johnson transformation: (a) Original image; (b) Image with gamma noise; (c) Residual noise; (d) Residual noise after transformation; (e) Original noise distribution; (f) Transformed noise distribution; (g) Noise distribution of stacking image patches; (h) Noise distribution of stacking transformed patches.}
	\label{fig_13}
\end{figure}

Real SAR images from multiple sensors such as Sentinel-1, Gaofen-3, and TerraSAR-X cover diverse scenes ranging from urban areas to complex natural terrains. These datasets often contain rich structural details, high-frequency textures, and numerous man-made targets, posing significant challenges for despeckling algorithms. On these datasets, we evaluated and compared the performance of multiple methods. Since real SAR images lack noise-free references, we used ENL, EPI, EPD, SQI, and mean intensity as evaluation metrics. Here, ENL indicates the degree of smoothing, while EPI and EPD measure edge preservation. As ENL often rewards excessive smoothing and EPI/EPD focus on preserving edge structures—which may result in insufficient noise suppression—it is crucial to jointly consider these metrics to obtain an objective and comprehensive evaluation.

%

\begin{table}[t]
	\caption{Quantitative Experiment on FARAD-X Samples (Mean = 10.05)\label{tb10}}
	\scriptsize
	\centering
	\newcolumntype{C}{>{\centering\arraybackslash}X}
	
	\begin{tabularx}{\linewidth}{>{\raggedright\arraybackslash}p{1.6cm}CCCCCC}
		\toprule
		\textbf{Method} & \textbf{ENL} & \textbf{EPI} & \textbf{EPD(H)} & \textbf{EPD(V)} & \textbf{SQI} & \textbf{Mean} \\
		\midrule
		ANLM\cite{xiao2020asymptotic}    & 0.765 & 0.343 & 15.81 & 22.66 & 0.658 & 9.58 \\
		DIP \cite{ulyanov2018deep} & 0.988 & 0.015 & 2.97  & 6.77  & 0.477 & 13.65 \\
		DnCNN\cite{zhang2017beyond}   & 0.847 & 0.330 & 13.01 & 19.58 & 0.703 & \underline{10.24} \\
		SARCAM\cite{9633208}  & 0.952 & 0.080 & 5.88 & 8.13 & 0.752 & 9.46 \\
		SAR2SAR\cite{dalsasso2021sar2sar} & \textbf{2.343} & 0.052 & 2.64 & 3.66 & \textbf{0.888} & 14.93 \\
		AGSDNet\cite{thakur2022agsdnet} & 0.668 & \textbf{0.887} & \textbf{54.97} & \textbf{63.48} & 0.606 & \underline{9.86} \\
		SIFSDNet\cite{thakur2022sifsdnet} & 1.056 & 0.104 & 5.96 & 8.56 & \underline{0.807} & 10.54 \\
		MONet\cite{vitale2023sar}   & 0.929 & 0.261 & 9.78 & 14.27 & 0.746 & 11.05 \\
		RDDPM \cite{hu2024sar}  & 0.379 & 0.389 & 19.79 & 23.28 & 0.406 & 7.94 \\
		CL-SAR \cite{fang2024contrastive} & 0.685 & 0.111 & \underline{51.32} & \underline{62.27} & 0.445 & 10.65 \\
		S3DIP \cite{albisani2025self} & 1.232 & 0.008 & 1.26  & 2.08  & 0.512 & 20.61 \\
		Ours    & \underline{1.256} & \underline{0.430} & 20.77 & 27.88 & 0.755 & \textbf{10.07} \\
		\bottomrule
	\end{tabularx}
\end{table}

\begin{table}[t]
	\caption{Quantitative Experiment on FARAD-Ka Samples (Mean = 9.63)\label{tb11}}
	\scriptsize
	\centering
	\newcolumntype{C}{>{\centering\arraybackslash}X}
	
	\begin{tabularx}{\linewidth}{>{\raggedright\arraybackslash}p{1.6cm}CCCCCC}
		\toprule
		\textbf{Method} & \textbf{ENL} & \textbf{EPI} & \textbf{EPD(H)} & \textbf{EPD(V)} & \textbf{SQI} & \textbf{Mean} \\
		\midrule
		ANLM\cite{xiao2020asymptotic}    & 1.151 & 0.305 & 15.54 & 22.56 & 0.803 & 9.21 \\
		DIP \cite{ulyanov2018deep} & 1.584 & 0.028 & 7.19  & 13.84 & 0.503 & 14.41 \\
		DnCNN\cite{zhang2017beyond}   & 1.225 & 0.296 & 13.02 & 19.53 & 0.837 & 9.82 \\
		SARCAM\cite{9633208}  & 1.099 & 0.359 & 17.33 & 24.27 & 0.773 & 9.28 \\
		SAR2SAR\cite{dalsasso2021sar2sar} & \textbf{2.061} & 0.081 & 3.64 & 5.46 & \textbf{0.928} & 10.99 \\
		AGSDNet\cite{thakur2022agsdnet} & 0.967 & \textbf{0.680} & \underline{35.66} & \textbf{43.35} & 0.724 & \underline{9.46} \\
		SIFSDNet\cite{thakur2022sifsdnet} & 1.199 & 0.520 & 22.18 & 29.02 & 0.827 & 10.09 \\
		MONet\cite{vitale2023sar}   & \underline{1.382} & 0.312 & 11.98 & 17.77 & 0.887 & 10.47 \\
		RDDPM \cite{hu2024sar}  & 0.563 & 0.360 & 23.90 & 27.14 & 0.449 & 7.41 \\
		CL-SAR \cite{fang2024contrastive} & 0.939 & 0.097 & 34.99 & 36.80 & 0.462 & 9.94 \\
		S3DIP \cite{albisani2025self} & 1.524 & 0.026 & 3.05  & 5.63  & 0.530 & 20.46 \\
		Ours    & 1.361 & \underline{0.541} & \textbf{36.47} & \underline{41.35} & \underline{0.892} & \textbf{9.61} \\
		\bottomrule
	\end{tabularx}
\end{table}

%

Figure \ref{fig_6}, \ref{fig_7}, and \ref{fig_8} illustrates the despeckling results on Sentinel-1, Gaofen-3, and TerraSAR-X data. Our method demonstrates strong generalization capability, effectively suppressing speckle noise while preserving fine structural details. Unlike traditional methods that often produce over-smoothed or blurred edges, our approach maintains linear features, building contours, and texture variations more faithfully. Compared with deep learning-based methods, the sparse representation model can adapt dynamically without being limited by out-of-distribution data or prone to overfitting. The despeckling results for Sentinel-1, Gaofen-3, and TerraSAR-X images consistently show clear edges, sufficient noise suppression, and good visual consistency across various scenes.

\begin{figure}[t]
	\centering
	\includegraphics[width=\linewidth]{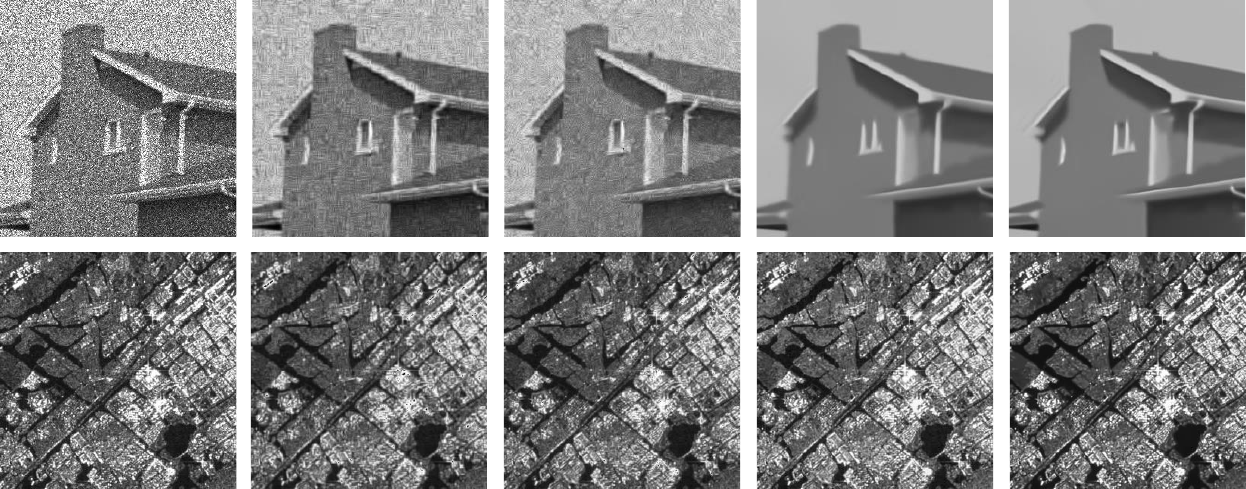}
	\caption{Ablation study, column from right to left: noisy image, denoised image (without Log-Yeo-Johnson transformation and auxiliary matrix), denoised image (without auxiliary matrix), denoised image (without Log-Yeo-Johnson transformation) and our proposed method.}
	\label{fig_14}
\end{figure}

\begin{table}[t]
	\scriptsize
	\centering
	\caption{Ablation studies. Evaluate the model performance of different module configurations. The symbol "\checkmark" means used in the model, and the symbol "\texttimes" means not used. \label{tb12}}
	\begin{tabularx}{\linewidth}{cccccccc}
		\toprule[1pt]
		\multirow{2}{*}{} & \multirow{2}{*}{\textbf{$w_1 \& w_2$}} & \multirow{2}{*}{\textbf{Transform}} & \multicolumn{2}{c}{\textbf{Synthetic(4-looks)}} & \multicolumn{3}{c}{\textbf{Real}}  \\
		& & & \multicolumn{1}{c}{\textbf{PSNR}} & \multicolumn{1}{c}{\textbf{SSIM}} & \multicolumn{1}{c}{\textbf{ENL}} & \multicolumn{1}{c}{\textbf{EPI}} & \multicolumn{1}{c}{\textbf{EPD}} \\ 
		\midrule
		(a) & \texttimes & \texttimes & 22.62 & 65.94 & 2.486 & 0.571 & 9.85 \\
		(b) & \texttimes & \checkmark & 23.21 & 67.21 & 2.537 & 0.599 & 10.32 \\
		(c) & \checkmark & \texttimes & \underline{25.60} & \underline{74.79} & \textbf{2.861} & \underline{0.658} & \underline{13.78} \\
		(d) & \checkmark & \checkmark & \textbf{25.98} & \textbf{77.13} & \underline{2.784} & \textbf{0.733} & \textbf{14.33} \\
		\bottomrule[1pt]
	\end{tabularx}
\end{table}

From the quantitative perspective, Tables \ref{tb6}, \ref{tb7}, and \ref{tb8} show that our method achieves superior overall performance on key metrics compared to other existing methods, demonstrating stronger capability to preserve structural and edge information. Moreover, our approach effectively maintains the mean intensity of the images, with the mean values significantly outperforming those of other methods across all datasets. Although the ENL metric of our method is slightly lower than that of some methods, this is because excessive smoothing can artificially raise ENL at the expense of detail loss. In contrast, our method effectively controls residual noise while retaining as much original detail and texture as possible, achieving a good balance among multiple metrics. It can be observed that certain methods, such as RDDPM, show clear signs of overfitting, as they perform well on Sentinel-1 but fail to deliver satisfactory results on other datasets. SAR2SAR, SARtrans, and MONet tend to cause over-smoothing in lower-resolution satellite scenarios, while AGSDNet and SIFSDNet struggle to remove noise efficiently, resulting in lower ENL but higher edge-preservation metrics.

For higher-resolution scenarios, we further verified the effectiveness of our method on miniSAR and FARAD datasets. These high-resolution datasets contain finer structural patterns, subtle texture variations, and sharper edges, where improper processing by traditional sparse representation or aggressive denoising methods can easily lead to detail loss or block artifacts. By introducing a non-local sparse representation, our model fully exploits the non-local self-similarity within images, enabling effective noise suppression and detail preservation under high-resolution conditions. As shown in Figures \ref{fig_9}, \ref{fig_10}, and \ref{fig_11}, our method achieves despeckling on miniSAR and FARAD without introducing noticeable blocky artifacts or excessive smoothing. Tables \ref{tb9}, \ref{tb10}, and \ref{tb11} provide the quantitative evaluation results, demonstrating that the sparse representation-based method consistently delivers the best comprehensive performance, effectively removing speckle noise while preserving edge structures and texture information.

The proposed method’s flexibility in adaptive patch grouping and iterative refinement further enhances its ability to handle varying backscatter intensities and complex scenarios. This adaptability enables the method to be widely applicable to various SAR systems and real-world applications. In summary, the experimental results on Sentinel-1, Gaofen-3, TerraSAR-X, miniSAR, and FARAD datasets clearly show that our approach achieves competitive despeckling performance across different resolutions and sensor types.

\begin{figure*}[t]
	\centering
	\includegraphics[width=0.86\textwidth]{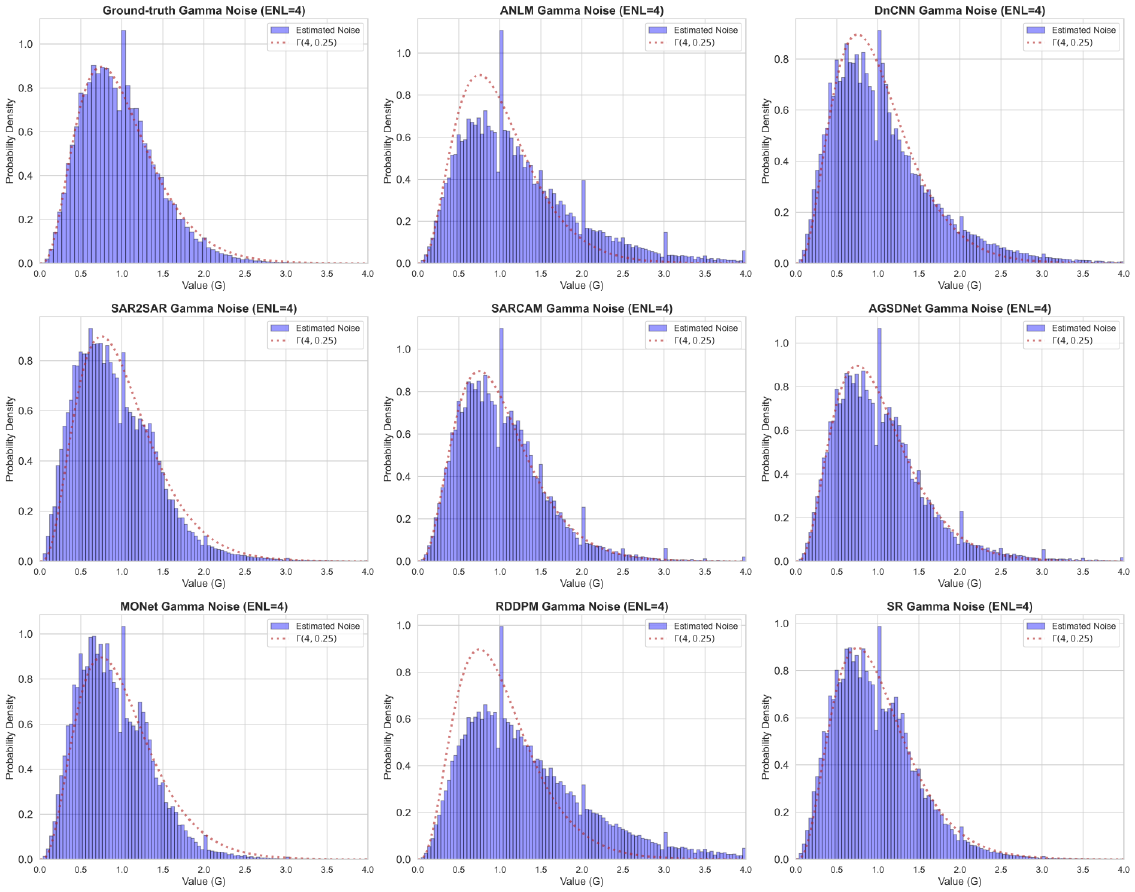}
	\caption{Noise distribution comparison before and after denoising: histograms represent the removed noise distribution, and red dashed lines denote the ground-truth Gamma distribution.}
	\label{fig_12}
\end{figure*}

\subsection{Ablation Study}

In this section, we validate the effectiveness of the Log-Yeo-Johnson transformation and two auxiliary matrices. Experiments were conducted on synthetic images to examine the residual noise before and after transformation, with histograms provided to illustrate noise distribution. Figure \ref{fig_13} presents the results of ablation study. Observing the histograms, the post-transformation noise distribution in Figure \ref{fig_13}(f) is significantly closer to a Gaussian distribution compared to the pre-transformation noise in Figure \ref{fig_13}(e). Similarly, the noise distribution of stacked image patches after transformation (Figure \ref{fig_13}(h)) shows a notable improvement in Gaussianity compared to the pre-transformation distribution in Figure \ref{fig_13}(g). These results demonstrate the effectiveness of our proposed method. Figure \ref{fig_14} visually illustrates the results of ablation study. Introducing the two auxiliary matrices significantly reduces image artifacts and improves the quality of the denoised image, while Log-Yeo-Johnson transformation enables the model to better preserve image details. Table \ref{tb12} provides quantitative results of the ablation study, which shows that two auxiliary matrix play important roles in this approach.

\subsection{Validation of Noise Characteristic}


As mentioned before, the observed intensity $y$ can be modeled by a multiplicative model in SAR imaging:

\begin{equation*}
	y = x \cdot n
\end{equation*} where $y$ denotes the underlying noise-free backscatter, and $n$ represents the speckle noise field. For fully developed speckle, $n$ follows a Gamma distribution:

\begin{equation}
	n \sim \Gamma(k, \theta)
\end{equation} where $k$ is the shape parameter and $\theta$ is the scale parameter. For an image with an Equivalent Number of Looks equal to $L$, the parameters satisfy:

\begin{equation*}
	k = L, \quad \theta = \frac{1}{L}
\end{equation*}

The probability density function is then:

\begin{equation*}
	p(n)=\frac{L^L n^{L-1} e^{-L n}}{\Gamma(L)}
\end{equation*}


Given a despeckled image $x$ and the corresponding noisy observation $y$, the residual noise field can be estimated as:

\begin{equation}
	G_{i,j} = 
	\frac{y_{i,j}}{x_{i,j}},
	\quad \text{for} \quad x_{i,j} > 0
\end{equation}

By aggregating all valid pixel-wise ratios, one obtains a sample set $\{ G_n \}$


To verify whether the residual noise follows the assumed speckle model, the empirical distribution of $\{ G_n \}$ is fitted to a Gamma distribution using maximum likelihood estimation (MLE). The fitted parameters $\hat{k}$ and $\hat{\theta}$ are estimated by:

\begin{equation}
	(\hat{k}, \hat{\theta}) 
	= \arg\max_{k,\,\theta} \prod_{n} p(G_n; k, \theta)
\end{equation}

The fitted shape $\hat{k}$ and scale $\hat{\theta}$ are then compared to the theoretical values: $k = L,$ $\theta = 1/L$.

A good despeckling algorithm should produce a residual noise field whose statistical distribution is consistent with the expected model. As shown in figure \ref{fig_12} and table \ref{tb5}, our proposed method achieves the best overall performance among all compared methods. Its estimated Mean value (0.9816) is the closest to the ideal value of 1, indicating that the radiometric information is well preserved. Meanwhile, its $\theta$ (0.2574) is also the closest to the theoretical value (0.25 for ENL = 4), demonstrating excellent consistency with the speckle noise model. Compared to other methods, our proposed approach achieves a better balance between noise suppression and detail preservation, highlighting its effectiveness for removing gamma noise.

\begin{table}[t]
	\caption{Consistency of residual with ground-truth Gamma distribution (BEST VALUES IN BOLD)}\label{tb5}
	\centering
	\scriptsize
	\newcolumntype{C}{>{\centering\arraybackslash}X}
	\begin{tabularx}{\linewidth}{p{1.0cm}CCCCC}
		\toprule
		Method & ANLM & DnCNN & SAR2SAR & SARCAM & SARtrans \\
		\midrule
		Mean & 1.2741 & 1.0629 & 0.9299 & 1.0424 & 0.8829 \\
		$\theta$ & 0.7681 & 0.4090 & \underline{0.2657} & 0.3427 & 0.2729 \\
		\bottomrule
	\end{tabularx}
	
	\vspace{1mm}
	
	\begin{tabularx}{\linewidth}{p{1.0cm}CCCCC}
		\toprule
		Method & AGSDNet & SIFSDNet & MONet & RDDPM & SR \\
		\midrule
		Mean & 1.0397 & \underline{0.9727} & 0.9101 & 1.5516 & \textbf{0.9816} \\
		$\theta$ & 0.3766 & 0.2653 & 0.2039 & 4.3852 & \textbf{0.2574} \\
		\bottomrule
	\end{tabularx}
\end{table}


%

\section{Discussion}

\subsection{Integration of Noise Transformation \& Non-local Sparsity}

This paper is inspired by \cite{ma2024despeckling}, where the gamma noise is transformed into approximate Gaussian noise. We here introduce non-local sparsity to effectively consider image similarity and transform the independent samples of Gaussian noise into statistically meaningful Gaussian distribution, satisfying the noise prior assumption of the model. As shown in Figure \ref{fig_13}, it can be observed that the noise in independent samples does not necessarily follow a Gaussian distribution. Stacked image blocks can better approximate a Gaussian distribution.

\subsection{Physical Meaning of $w_1$ and $w_2$}

This paper introduces two adaptive weighting matrices, $w_1$ and $w_2$, into sparse representation modeling to more effectively suppress the spatially inhomogeneous speckle noise in SAR images while preserving critical structural information. Specifically, $w_1$ adaptively weights the residual term based on the local noise level in different regions. In areas with higher noise levels, the fitting constraint is appropriately relaxed (this constraint is not enforced) during optimization, which helps prevent overfitting to highly noisy patches and significantly suppresses pseudo-structures. Conversely, regions with lower noise levels are subject to stronger constraints, enabling better preservation of fine details.

Meanwhile, $w_2$ incorporates a Laplacian prior on the sparse coefficients to adaptively adjust the importance of each dictionary atom in sparse regularization. This mechanism reduces the sparsity penalty on key features while increasing it for irrelevant noise components, thereby further enhancing the robustness and denoising performance of the sparse representation. This dual adaptive modeling, which accounts for both the statistical characteristics of noise and the structural features of the dictionary, substantially improves the proposed method’s ability to remove speckle noise and preserve details under varying noise levels and complex texture scenarios.

\subsection{Limitations and Future Work}

While our proposed method demonstrates strong generalization and competitive performance across a variety of SAR datasets, some limitations remain. First, although the Log-Yeo-Johnson transformation improves the Gaussianity of noise, the approximation may still be insufficient in regions with extremely heterogeneous textures or under very low ENL conditions. This can lead to suboptimal performance in highly non-Gaussian environments. Second, our proposed method requires only a single sparsity parameter to achieve adaptive despeckling. This hyperparameter was fixed throughout the experiments, which led to slight oversmoothing in certain scenarios, such as high-resolution SAR despeckling.

In future work, we plan to explore learnable noise transformations that adaptively approximate noise characteristics based on data distribution, potentially leveraging lightweight neural modules. Moreover, integrating semantic information or leveraging hybrid frameworks combining sparse coding and deep priors could further improve robustness and adaptivity.

\section{Conclusion}

This paper introduces a simple and intuitive SAR despeckling method that combines Log-Yeo-Johnson transformation with compressive sensing theory to address the challenges of non-Gaussian despeckling in SAR images. By leveraging non-local sparse representation and auxiliary matrices, the proposed approach enhances noise characterization and sparsity, yielding superior despeckling performance. Experiments on synthetic and real images demonstrate method's effectiveness in noise suppression and detail preservation.

\bibliographystyle{IEEEtran}
\bibliography{reference}

\vfill

\end{document}